\documentclass[twocolumn]{aastex63}

\usepackage{wrapfig}
\usepackage{tcolorbox}
\usepackage{rotating}
\usepackage[utf8]{inputenc}
\usepackage{amsmath}
\usepackage{booktabs}
\usepackage{floatflt}
\usepackage{lineno}

\begin{document}

\title{Thermal and Optical Characterization of Near-Earth Objects: Science Commissioning of the Recently Upgraded Mid-Infrared Camera MIRSI on the NASA Infrared Telescope Facility}

\correspondingauthor{Andy J. L\'{o}pez-Oquendo}
\email{al2987@nau.edu}

\author[0000-0002-2601-6954]{Andy J. L\'{o}pez-Oquendo}
\affiliation{NASA Goddard Space Flight Center, Solar System Exploration Division, Greenbelt, MD 20771, USA}
\affiliation{Department of Astronomy and Planetary Science, Northern Arizona University, Flagstaff, AZ 86011, USA}
\affiliation{Visiting Astronomer at the Infrared Telescope Facility, University of Hawaii, 640 N. Aohoku Pl., Hilo, HI, USA}

\author[0000-0002-5599-4650]{Joseph L. Hora}
\affiliation{Center for Astrophysics $|$  Harvard {\&} Smithsonian, 60 Garden St., MS-65, Cambridge, MA 02138, USA}

\author[0000-0003-4580-3790]{David E. Trilling}
\affiliation{Department of Astronomy and Planetary Science, Northern Arizona University, Flagstaff, AZ 86011, USA}

\author{Howard A. Smith}
\affiliation{Center for Astrophysics $|$  Harvard {\&} Smithsonian, 60 Garden St., MS-65, Cambridge, MA 02138, USA}

\begin{abstract}

Mid-infrared (mid-IR) observations of Near-Earth Objects (NEOs) have historically been a valuable tool for understanding their physical properties. However, the current state of mid-IR instruments on ground-based telescopes places several limitations on performing thermal characterization of NEOs. The complexity of maintaining these instruments in operational conditions on telescopes has led to their decommissioning. Here, we present the first science commissioning observations out to 12.5~${\mu}$m from the upgraded Mid-Infrared Spectrograph and Imager (MIRSI) at the NASA-IRTF. We obtained 42 observations of 31 NEOs and derived their diameters and albedos. Since MIRSI allows simultaneous optical observations with its MIRSI Optical Camera (MOC), we were able to determine the absolute magnitude for most of the targets at the time of the thermal acquisition. We present ejecta characterization for the Didymos system from observations made 11 hours and 9 days after the Double Asteroid Redirection Test (DART) impact. We present albedo and size measurements for (98943) Torifune 2001 CC21, the fly-by target of the Japanese Extended Hayabusa2 Mission (Hayabusa2$\sharp$). We also highlight several applications that the MIRSI system will provide for future airless body characterization, such as constraining thermal inertia from simultaneous optical and thermal lightcurves. This work also demonstrates the importance of having MIRSI as an available rapid-response instrument for planetary defense purposes.

\end{abstract}

\keywords{NEOs, Small Solar System bodies, Mission Support, Asteroids: general --- techniques: photometry, thermal modeling}

\section{Introduction \label{sec:intro}}

The mid-infrared (mid-IR) region of the electromagnetic spectrum has been key for the characterization of airless bodies and, by extension, to the understanding of the formation and evolution of the Solar System. Particularly, significant improvements in the studies of Near-Earth Objects (NEOs), or those airless rocky bodies whose orbit brings them close to the Earth's vicinity, have been possible from ground- and/or space-based thermal observations \citep{DELBO2003116, Delbo2015, Trilling_2016, Mainzer2015}. NEOs represent a potential risk of impact to the Earth and, consequently, to human civilization. Thus, detecting and characterizing them is a matter of concern and a priority for planetary defense practices and implementation.

While the proximity of NEOs poses a potential threat to life on Earth, their orbits make them ideal targets for characterization upon close approaches. Determining the physical properties of NEOs is crucial for developing efficient mitigation strategies in the event of possible future impacts \citep{Farnocchia2015, Harris2015} or for planning a deflection effort \citep{Rivkin_2021, Rivkin2023, Raducan2024}. One of the primary objectives of planetary defense efforts is to characterize an object's size and surface properties, including composition, surface roughness, and bulk density \citep{Reddy2022}. Constraining the albedo of a potential impactor can aid in confirming its taxonomy \citep{Thomas2014} by cross-correlating results from various characterization techniques, as demonstrated in recent rapid-response planetary defense campaigns \citep{Reddy_2024}. This information is then used to establish a plausible range of bulk density and understand the severity of the impact \citep[i.e., distinguishing between high- vs. low-density objects;][]{CARRY201298}.

The current state of ground-based remote sensing characterization is hindered by the lack of instruments available to carry out NEOs observations at these wavelengths. Cryogenically cooled thermal instruments are challenging and increasingly expensive to maintain at operating temperatures for long periods of time. Due to their operational complexities and difficulties in mid-IR ground-based observing, many mid-IR instruments have been decommissioned. Examples include the MIRAC cameras \citep{Hoffmann_1994} used at the NASA Infrared Telescope Facility (IRTF) and the United Kingdom Infrared Telescope (UKIRT) from the late 1980's - 2000, the T-ReCS and Michelle instruments used at the Gemini Observatory \citep{DeBuizer_2005}, and the Mid-Infrared Spectrometer and Imager (MIRSI), which operated at the IRTF from the early 2000s \citep{Deutsch2003, Kassis2008} until 2011. 

Although the mid-IR is currently experiencing a fruitful era with the JWST Mid-InfraRed Instrument \citep[MIRI;][]{Rieke_2015, Wright2015}, for many purposes, it is necessary to have an ``easily'' accessible and operational, rapid-response thermal infrared camera with fewer pointing restrictions. For example, recently discovered asteroids require a fast observing response due to their short observability upon close approach \citep{GALACHE2015155, Lopez-Oquendo2023}. Furthermore, the absence of available programs, such as the Arecibo Planetary Radar System \citep{Virkki_2022}, for characterizing small asteroids approaching Earth underscores the need to explore alternative methods for size estimation. 

In light of the usefulness of thermal infrared instruments and in recognition of their contribution to NEOs characterization, in 2014 our team was funded by the NASA Solar System Observations/NEOO program to upgrade MIRSI's liquid cryogenic system to a closed-cycle cooler. In addition, the upgrade included the construction and delivery of an optical camera attached to the MIRSI system to allow for simultaneous thermal and optical observations. Such an upgrade would allow rapid observations of recently discovered NEOs, mission-support observations of targets of interest, and a high-quality diameter and albedo NEO survey of objects whose properties are unknown.

In this paper, we present the first MIRSI science commissioning results on airless body characterization from simultaneous optical and thermal observations acquired during the 2022 to 2024 period. We describe our NEO program, along with the data collection, reduction, and analysis. We highlight several potential uses of MIRSI as a key instrument for planetary defense, mission support, and additional planetary science applications. We present a detailed MIRSI overview, including the design and performance, in our parallel instrument paper \citep{Hora2024}.

\section{Data acquisition and reduction}

\subsection{Target selection \label{sec:observability}}

To select our targets, we built an observability \textit{Python} script to retrieve observable NEOs from the NASA IRTF. We downloaded the up-to-date list of discovered NEOs from the Minor Planet Center (MPC)\footnote{\url{https://minorplanetcenter.net/data}} and utilized the JPL-Horizons System\footnote{\url{https://ssd.jpl.nasa.gov/horizons/}} \citep{Giorgini_2015} to predict their observability. We explored the NEO's brightness through the time span of an IRTF observing semester to determine when the NEO reached its brightest point. NEOs with airmasses $>$2.0 and V-band magnitudes $>$20~mags and whose visibility window was shorter than 30 minutes were removed from the candidacy list. We implemented the Near-Earth Asteroid Thermal Model \citep[NEATM;][]{Harris1998} to our observability script to predict the NEO SEDs using the object's geometry at the time of their maximum visible brightness (see \S \ref{sec:TherMod} for more discussion about NEATM). For the SED calculation, we assumed an average visible albedo $p_{V}$ of 0.15, a beaming parameter of 1, a thermal emissivity of 95\%, and the absolute magnitude from the JPL-Horizons system. Then, the flux at 10.5~${\mu}$m was retrieved from the SED, and required integration times to achieve a 10$\sigma$ detection were made based on MIRSI's performance \citep{Hora2024}. 

\begin{figure}[!]
    \centering
    \includegraphics[width=0.47\textwidth]{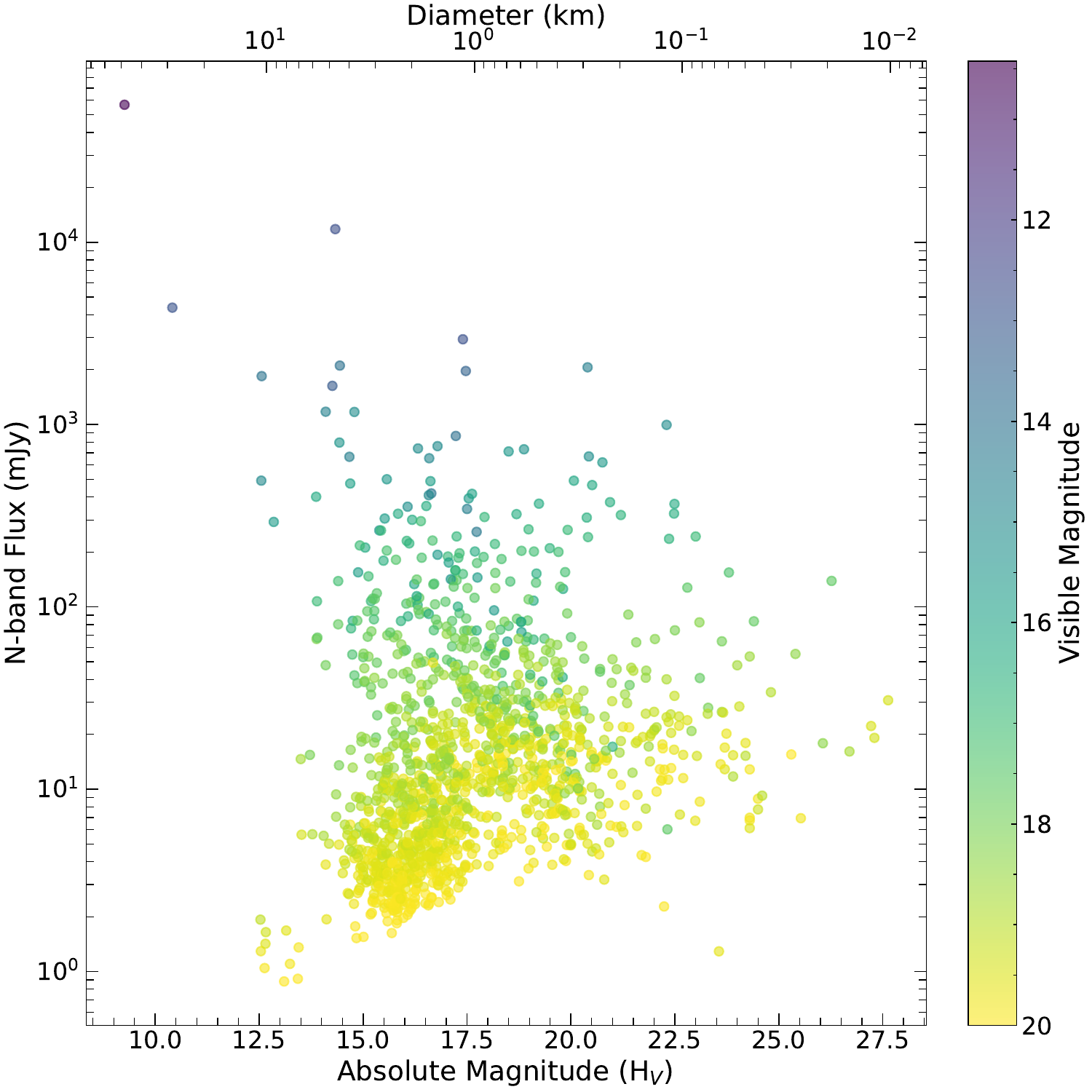}
    \caption{Predicted N-band flux as a function of Solar System absolute magnitude $H_{V}$ for all observable NEOs from the NASA IRTF during a given observing semester. The top axis is the estimated diameter from $H_{V}$ assuming a 15\% albedo in the \citet{Harris2002} relationship. Each circle's color indicates the visible magnitude of the NEO at the time it reached peak in brightness. Yellow represents fainter objects (larger visible magnitude), while purple represents brighter ones.}
    \label{fig:predictedflux}
\end{figure}

In Figure~\ref{fig:predictedflux}, we show the predicted N-band flux as a function of absolute magnitude for every observable NEO from the NASA IRTF for a typical semester (i.e., from February through July). As illustrated in Figure~\ref{fig:predictedflux}, NEO fluxes can range from a few milli-Jansky to a few Jansky (under the condition of a constant albedo assumption). From such observable NEOs, those with N-band fluxes below 50~mJy were discarded. Finally, our strategy consisted of selecting objects whose predicted flux was sufficiently (i.e., $>$200~mJy) high to be detected in a reasonable 3 to 5~hour telescope block as based on MIRSI's performance \citep{Hora2024} at the time of our observations. By implementing this selection, we combined different candidates per observing block to maximize NEOs observations.

As part of our observational campaign strategy, we intentionally aimed to observe objects that had previous diameters and albedos determined either from ground or space-based telescope observations to compare and contrast with. We also targeted specific objects with the purpose of mission support observations. 
In addition, to highlight MIRSI's capability for NEO studies and planetary defense, we also carried out rapid-response observations of recently discovered NEOs. Since these objects are confirmed near their proximity to Earth and fade quickly after a close approach, their observability prediction can only be made within a few days or hours of a scheduled observing block. Thus, when available and bright enough, we aimed to observe them. Otherwise, they were observed as Targets of Opportunity.

\subsection{Observations \label{sec:observation}}
Our observations were conducted using the MIRSI camera at the 3-meter NASA Infrared Telescope Facility (IRTF) located at Mauna Kea, Hawaii. MIRSI has the capability to perform broadband and narrowband observations at imaging mode and low-resolution grism mode spectra within the 8-14 and 17-26~$\mu$m atmospheric windows. In imaging mode, the system has a field of view of 86$^{\prime\prime}$$\times$63$^{\prime\prime}$ with a plate scale of 0\farcs267~pixel$^{-1} \times 0\farcs264 $~pixel$^{-1}$. MIRSI's system possesses a dichroic mirror mounted at 45$^{\circ}$, which enables the infrared (IR) light from a science target to be collected by MIRSI while the optical light is directed to the MIRSI Optical Camera (MOC). MOC is a fast readout CCD camera whose operations mirror those of MORIS \citep{Gulbis2010, Gulbis2011}, the optical camera of the SpeX instrument at the IRTF \citep{Rayner2003}. Having the capability of acquiring simultaneous optical and IR data provides an incredible advantage for NEO studies. MOC allows optical guiding on-source, providing highly accurate position of objects in the MIRSI field-of-view, subsequently enabling blind stacking of images for objects not detected in a single MIRSI exposure (i.e., targets with thermal emissions $<$1~Jy). In addition, the ability to synchronously collect visible data while performing an IR observation allows for the target's optical photometry characterization (i.e., rotation period, color, lightcurve amplitude, and absolute magnitude). 

Our observations were performed using MIRSI's imaging mode. Because most of the science observations presented here were carried out while MIRSI had an engineering IR array installed and because its sensitivity was degraded from the nominal science array, the camera's performance limited our range of observations setup. In addition, the dimmer nature of most NEOs' (see Figure~\ref{fig:predictedflux}) also contributed to the observing strategy. For such reasons, we decided to sacrifice multiple spectral energy distribution (SED) data points, which would have required an extensive amount of telescope time at the expense of enhancing single-band detections of NEOs. Thus, most of our targets were observed utilizing the broad 10.5~$\mu$m filter (N-band). The broad N-band filter offers better sensitivity compared to others \citep{Hora2015} and captures most of the NEO SED maxima, making this region ideal for their observations. 

The MIRSI IR data were obtained using the telescope nod and offsets to obtain images of the NEO at different locations on the array for sky background subtraction. For asteroid observations, we used integration times of 15~ms with cycles between 300 and 500 coadded images, as we found those cycles yielded the best S/N for the N-band.\footnote{\url{https://irtfweb.ifa.hawaii.edu/~mirsi/}} Each cycle takes about 35~sec due to overheads in the array readout. To calibrate the NEO counts into physical flux units, we observed IR standard stars from the \citet{Cohen1999} catalog. For standard star observations, we used integration times of 15~ms with cycles of 200 coadded images.

With the current system performance, MIRSI allows a 10$\sigma$ detection of a 1.3~Jy source in around 3~minutes of clock time (i.e., including the telescope overheads and readout of the array). Since thermal models are generally not accurate to better than 10\% \citep{Trilling_2016}, we established a minimum signal-to-noise-ratio (SNR) of 10 as an appropriate goal for our thermal photometry. Thus, we observed each object for at least a minimum integration time to meet such detection requirements with the system performance as of fall 2022. In many cases, we spent longer than the required minimum time on objects whose predicted flux was less well-constrained, given that we had to assume unknown surface properties such as the albedo (see \S \ref{sec:observability}) or due to variable sky conditions. We also carried out MIRSI observations with the narrowband 7.8, 8.7, 9.8, 11.7, and 12.5~${\mu}$m filters. The multiband observations were performed for specific science targets whose predicted thermal fluxes (see \S \ref{sec:observability}) would allow successful detection in a reasonable amount of telescope time. 

As we acquired IR data, we also collected MOC optical photometry using the SDSS-r broadband filter to constrain the NEO absolute magnitude ($H$). We also used MOC to guide the object in the IR field. MOC can reach an SNR$=$10 on a 21~mags (V-band) object in around 1 minute. We typically used an integration time of 3 seconds for each asteroid since most of our objects were brighter than 18.5~mags at V-band (see Table \ref{tab:Table1}). MOC is essentially a clone of MORIS, and its primary purpose is for guiding. Due to the relatively small field-of-view (60 arcsec $\times$ 60 arcsec) several issues are imposed for frame-to-frame photometric calibration since stars will appear infrequently within the asteroid frame. Because of the expected and observed lack of available background objects for calibration, we opted to individually observe an SDSS optical standard star right after the NEO observations at a similar airmass to the NEO. In Table \ref{tab:Table1}, we show the observational circumstances of our program, including the names of the optical and IR standard stars observed. We also show the NEO geometry information at the time of the NEO observation obtained from the JPL-Horizons system.

\subsection{Data reduction}
\subsubsection{MIRSI photometric reduction}
To process the thermal data, we used the updated MIRSI reduction pipeline\footnote{\url{https://github.com/jhora99/MIRSI}} \citep{Hora2024}. The MIRSI pipeline consists of a set of Python scripts that perform A-B subtracted images, pixel-to-pixel gain corrections, construct the mosaics, and perform aperture photometry, including airmass correction and flux calibration. We present a detailed description of the MIRSI pipeline in our parallel instrument paper \citep{Hora2024}. 

\begin{figure}[!]
    \centering
    \includegraphics[width=0.47\textwidth]{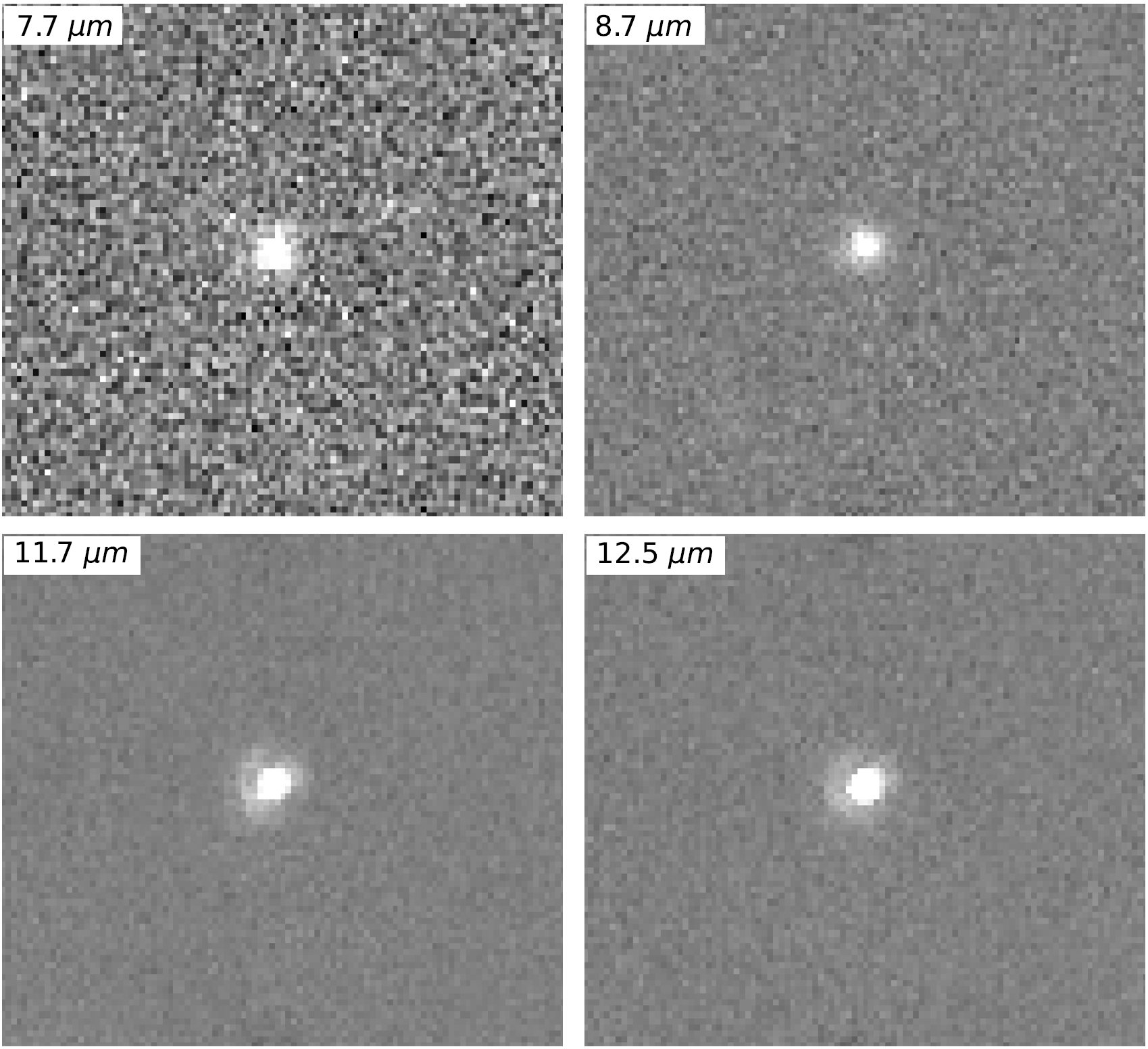}
    \caption{Mid-IR mosaics of NEO 7482 obtained with MIRSI's 7.7 (top-left), 8.7 (top-right), 11.7 (bottom-left), and 12.5~$\mu$m (bottom-right) filters. Each image size is about 26\farcs7 horizontally and 21\farcs4 vertically.}
    \label{fig:7482_mosaic}
\end{figure}

The MIRSI pipeline initiates the imaging process by differencing the A-B exposures to remove the sky background and array readout pattern. Column-wise and row-wise offsets in the frame are removed by forcing the medians of the columns and the rows to be the same. The frames are then multiplied by a flat field image to correct for pixel-to-pixel gain variations. Bad pixels are masked so they are not included in the mosaic construction. Subsequently, a target mosaic, as shown in Figure~\ref{fig:7482_mosaic}, is created by aligning the individual frames based on the commanded offsets, and a sigma-clipped mean value is calculated at each position. A similar procedure is employed to acquire and reduce the standard star data. After mosaics were created, the pipeline performed the photometry using aperture photometry for each object, using an aperture radius of 2\farcs16 which was chosen to include $\sim$95\% of the signal from a point source as determined from standard star measurements. Including more pixels would have reduced the effective S/N of the photometry. This aperture radius was used for all filters for both the standard star and NEO measurements. The fluxes are airmass corrected, and the standard star photometry is used to determine the Jy/ADU conversion factor, which is then applied to the NEO photometry. Finally, the NEO flux is measured using the same aperture photometry techniques and parameters, and 1$\sigma$ uncertainty is obtained from background noise. There also exist calibration uncertainties at the 10-15\% level due to the sky variations, errors in the pixel gain correction, and uncertainties in the absolute calibration of the standard stars.

\subsubsection{MOC photometric reduction \label{sec:moc_red}}
To reduce the MOC optical data, we developed a Python script to perform photometry. We initiate downloading MOC frames from the IRTF archive. Then, we ran the code on individual collected nights. The script registers images for a given filter and reads the MOC image header information, which is then copied to the default files of the Source Extractor \citep[SExtractor;][]{Bertin1996}. Then, SExtractor is called to identify all objects within MOC's field and perform aperture photometry. The output photometry files are then read to perform time series analyses. Because MOC operates asynchronously from the IR camera and its beam-switching commands, some frames are taken while the telescope is moving between nod positions, thus smearing the optical image and contaminating the automated extracted photometry. We removed those output files corresponding to the damaged frames from the photometry analysis. We also performed a visual inspection of the raw photometry data and removed suspicious frames due to clouds or interference from background stars throughout the observations. Finally, the extracted photometry of the NEO is calibrated using the respective observed standard star listed in Table~\ref{tab:Table1}. We found a calibration offset of 0.36~mags by averaging all the offset measurements.

\subsection{In-band flux correction}

\begin{figure}[!]
    \centering
    \includegraphics[width=0.47\textwidth]{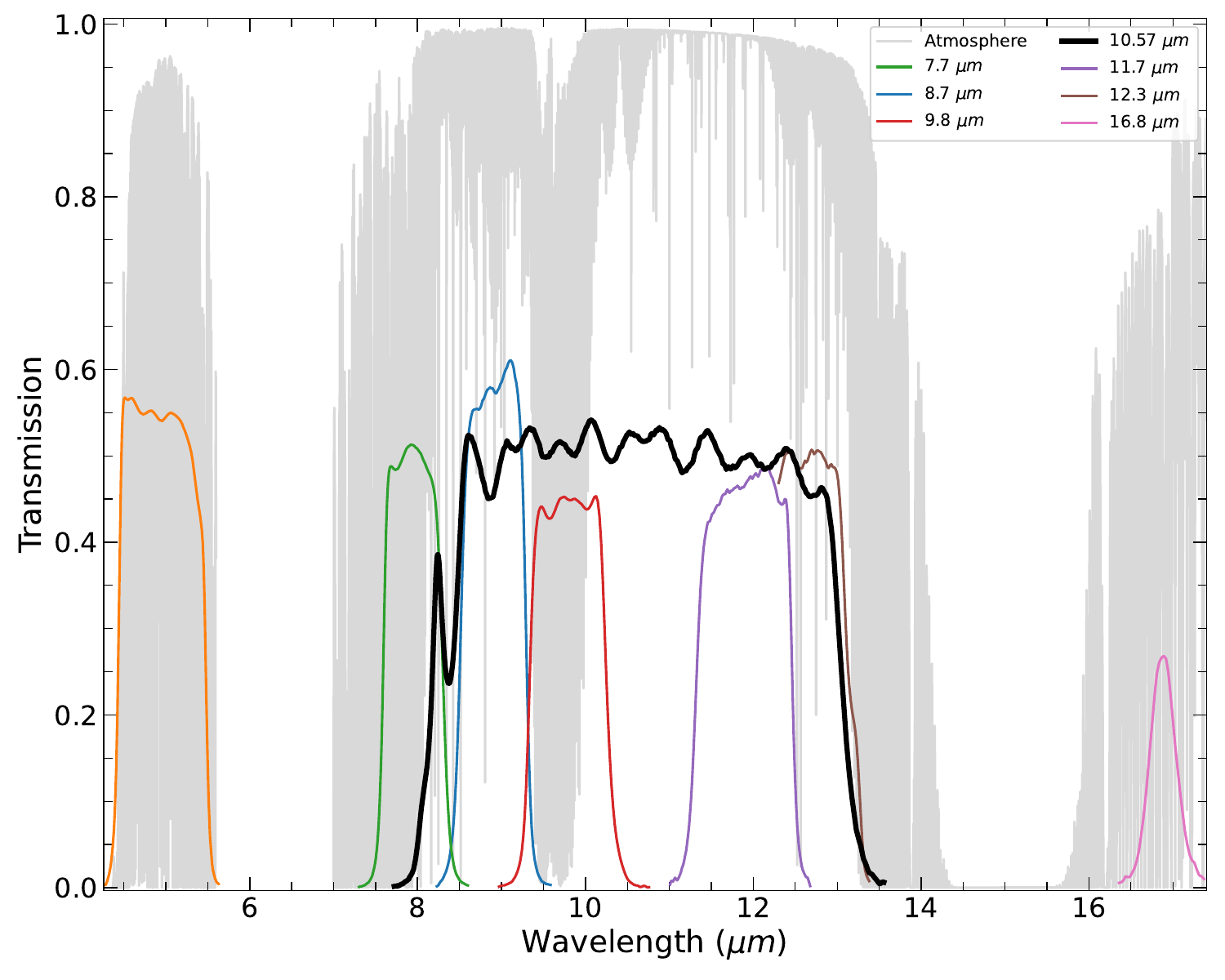}
    \caption{MIRSI transmission filter curves obtained at room temperature at the IRTF as a function of wavelengths in microns. Grey continuous curves correspond to the 1.6~mm atmospheric transmission from Mauna Kea.}
    \label{fig:filters}
\end{figure}

To properly calibrate the measured asteroid fluxes, color corrections induced by the system must be taken into account to account for the Vega system, as such, were implemented for the standard stars used here \citep{Cohen1999}. The terminology of color correction refers to the incorporation of isophotal wavelength and zero-magnitude flux to a source $F_{\lambda}$ whose spectrum differs from that of Vega $F^{Vega}_{\lambda}$, thus requiring a correction ($f$) to give the same signal as Vega \citep{Hanner1984, Tokunaga2005}, in this case, at the MIRSI's system bands. We incorporated the definition of zero-magnitude flux density and isophotal wavelengths described in \citet{Tokunaga2005} to compute the intrinsic flux of Vega (in $W~m^{-2}~{\mu}m^{-1}$) considering the MIRSI system (see Figure \ref{fig:filters}). We used the expression provided by \citet{Wright_2010} to fit the Vega continuum between the mid-IR wavelengths. We generated atmospheric transmission spectra using the ATRAN model \citep{Lord1992}. The atmospheric model was built by adopting the altitude of Mauna Kea, 1.6~mm of precipitable water, and an air mass of 1.2. We calculate S($\lambda$), the monochromatic in-band flux, as

\begin{equation}
 S(\lambda) = {\left( \frac{\int R(\lambda)A(\lambda)B(\lambda)\lambda d\lambda}{\int R(\lambda)\lambda d\lambda} \right)} 
 \label{eq:convolution}
\end{equation}
where B($\lambda$) is the asteroid SED, A($\lambda$) is the atmospheric transmission at Mauna Kea, R($\lambda$) is the filter response function for MIRSI (see Figure~\ref{fig:filters}), and $\lambda$ is the effective wavelength region of the filter response. For the asteroid thermal spectrum, we assumed a visible albedo $p_{V}$ of 0.15 and the geometry conditions at the time of the observation (see Table \ref{tab:Table1}).

We obtained a zeropoint magnitude on the Vega system of 9.283${\times}10^{-13}~W~m^{-2}~{\mu}m^{-1}$ or 34.597~Jy by using $F_{\nu} = ({\lambda}^{2}_{iso}/c)F_{\lambda}$ conversion relationship. There is an additional $\pm$1.5\% in the estimated zeropoint flux due to systematic uncertainty from the Vega spectrum \citep{Cohen1992}. We find that correction factors ranged between 0.9 and 1.2 and averaged around 0.97 for the NEOs observed here at the N-band; ultimately, each measured flux was divided by each correction $f$.

MIRSI's filter response functions, shown in Figure \ref{fig:filters}, were obtained at room temperature and not at the nominal working temperature of 5~K (-268.73~$^{\circ}$C). Such temperature differences would induce a wavelength shift in the response filters, thus causing the correction factors to vary as well. Quality tests on these filters by several companies have found that the wavelength shifts by around 1\% when measuring filter response functions at room compared to cold temperature (-200~$^{\circ}$C). To explore the degree of alteration in the measurements of the correction factors, we assumed the extreme case of a 5\% wavelength shift ($\Delta\lambda$). {We found that adding a 5\% $\Delta\lambda$ shift in the N-band filter wavelengths causes the correction factors to differ by 0.2\% from those when using the response function obtained at room temperature.}

\section{Thermal Modeling \label{sec:TherMod}}
In order to interpret the measured thermal fluxes by MIRSI, we utilize the well-tested Near-Earth Asteroid Thermal Model \citep[NEATM;][]{Harris1998}. The NEATM calculates the target's SED by computing the temperature distribution and assuming an energy balance between insolation and thermal emission and surface properties such as roughness. Subsequently, the surface radiation for those temperatures is calculated by integrating the Planck function over the visible hemisphere of a sphere. The modeled SED provided by NEATM, along with the combination of optical photometry ($H_{V}$), have been widely used to independently derive the diameter (D) and V-band albedo ($p_{V}$) of the planetary body \citep{DELBO2003116, Trilling_2010, Mainzer2011, Mueller_2011, Trilling_2016}. 

Previously, the ExploreNEOs program \citep{Harris_2011, Trilling_2016} found that single-band fits by the NEATM approach could yield diameters and albedos accurate within 20\% and 50\%, respectively \citep{Harris_2011}. Here, we utilized a similar approach as in those programs to fit our single N-band measurements to constrain the diameter of a body and, subsequently, the albedo with the simultaneously MOC-derived $H_{V}$ (see \S \ref{sec:Hmoc}) and the JPL $H_{V}$ solution.

\subsection{MOC Photometry Analysis \label{sec:Hmoc}}

Absolute magnitudes $H_{V}$ are usually poorly constrained for NEOs, often yielding inaccurate albedo estimations \citep{Gustafsson_2019} or high uncertainties in the solutions \citep{Trilling_2016, Masiero2021}. As described in \S \ref{sec:observation} and \ref{sec:moc_red}, the simultaneously collected optical data by MOC is crucial for NEOs studies since $H_{V}$ can be determined at the time of the thermal measurements, thus minimizing uncertainties in the $p_{V}$ determination \citep{Masiero2021}.

We collected optical data using the SDSS-r filter. To transform our MOC SDSS-r magnitudes to V-band, we used a spectral convolution approach presented in \citet{Lopez-Oquendo_2022, Lopez-Oquendo_2024} to retrieve colors from classified asteroids spectra contained in the MITHNEOS program \citep{Binzel2019}. We retrieved all the S-, X-, C-complex, and V-type asteroids $V - r$ color using the response function of each of these filters and found correction values of 0.386, 0.323, 0.308, and 0.397~mags, respectively. For objects without taxonomy classification, we adopted the average $V - r$ between S- and C-complex asteroids of 0.346~mags. The correction value was then added to the MOC SDSS-r visible photometry to obtain the visible magnitude $V_{MOC}$.

We used the \citet{Bowell1989} absolute calibration phase-magnitude (H-G) system to determine the absolute magnitude $H_{V}$ by using the following relationships:

\begin{multline}
H_{V} = V_{MOC}~+~2.5 log_{10}((1-G_{V})\Phi_1~+~G_{V}\Phi_2)~\\-~2.5log_{10}(R{\Delta})     
\label{eq:hv}
\end{multline}
\begin{align*}
    \Phi{_1} &= exp\left(-3.33 tan^{0.63}\left(\frac{\alpha}{2} \right) \right)\\
    \Phi{_2} &= exp\left( -1.87 tan^{1.22} \left( \frac{\alpha}{2} \right) \right)
\end{align*}


\noindent where $G_{V}$ is the phase curve slope parameter, $\alpha$ is the solar phase angle, $R$ is the heliocentric distance, and $\Delta$ is the geocentric distance of the object at the time of the observation (see Table \ref{tab:Table1}). For the $G_{V}$ slope parameter, we assumed values of 0.15, 0.25, and 0.4 for C-, S-, and X-complex asteroids, respectively, based on those observed for similar object taxonomy \citep{VERES201534}. For those objects whose taxonomy was unknown, we used a $G_{V}$ parameter of 0.15 as it is a typical value among asteroids \citep{Robinson2024}.

\subsection{Monte Carlo Simulations}

Thermal modeling with NEATM uses parameters, which in some cases have to be assumed from observed bulk properties of NEOs. These parameters are known to be significant contributors to the estimation of accurate diameters and albedos from thermal modeling. For instance, the beaming parameter ($\eta$) attempts to correct for a range of physical properties of the regolith (i.e., surface roughness and thermal inertia). The poorly constrained beaming parameter ${\eta}$ is known to be the main driver in uncertainties in derived diameters \citep{Trilling_2016}. In principle, ${\eta}$ can be constrained from thermal modeling if several data points of the SED exist. However, for single-band measurement as mostly performed here, this parameter has been obtained from empirical linear relationships between $\eta$ and solar phase angle $\alpha$ \citep{Wolters2008, Mainzer2011, Trilling_2016}. Here, we adopted the ${\eta}-{\alpha}$ relationship by \citet{Mainzer2011} to derive ${\eta}$ based on the object solar phase angle $\alpha$: ${\eta} = (0.00963 {\pm} 0.00015){\alpha}~+~0.761 {\pm} 0.009$. In our multiple-band thermal analyses, we use $\eta$ as a free-floating fitting parameter. For the thermal modeling analysis of small recently discovered asteroids, we implemented a Fast Rotator Model (FRM) due to the likely rapid rotation of these asteroids \citep{PRAVEC200012}. The FRM or the ``isothermal latitude model'' significantly varies from NEATM since the temperature distribution is only a function of the latitude. In other words, the temperature distribution on the surface of the body consists of isothermal latitudes with temperature peak at the equator and falling to zero towards the poles.

In order to obtain diameter and albedo estimates, we incorporated a Monte Carlo analysis with 10$^{6}$ trials to explore the range of solutions induced by various sources of uncertainties in thermal modeling with NEATM. To deal with the uncertainties in diameter, we assumed a 1$\sigma$ random Gaussian distribution around the nominal $\eta$ value derived from the \citet{Mainzer2011} linear relationship. While uncertainties in $H_{V}$ are known to cause negligible impacts in the diameter \citep{Mueller_2011}, they dominate uncertainties in the albedo $p_{V}$ \citep{Masiero2021}. To investigate the uncertainties in albedo, we assumed a random Gaussian distribution around the MOC photometry uncertainty. Here, we also present albedo solutions using the JPL-$H_{V}$. {For the thermal models with JPL-$H_{V}$, we assumed a random Gaussian distribution around ${\Delta}H=0.3~$mags \citep{PRAVEC2012365} as this has been a common practice on previous works \citep{Trilling_2016}}. We also included a 10\% uncertainty in the measured flux to account for systematic calibration uncertainty. Finally, the Monte Carlo simulations incorporated these parameters into the thermal modeling and used the ${\chi}^{2}$ minimization procedure. We then determined the likelihood probability function and retrieved the 16, 50, and 84 percentile values to obtain the best-fit solutions for diameter and albedo and their 1$\sigma$ uncertainties. 

To analyze the multiple band measurements, we used a Markov-Chain Monte Carlo \citep[MCMC;][]{Foreman2013} analysis to explore the diameter, $p_{V}$, and $\eta$ best-fit solutions. The MCMC simulations were executed by using 1000 walkers and 3000 steps to find the best-fit solutions. We used a similar statistical approach as mentioned before to retrieve the 16, 50, and 84 percentile values from the likelihood probability function.

\begin{figure}[h!]
    \centering
    \includegraphics[width=0.47\textwidth]{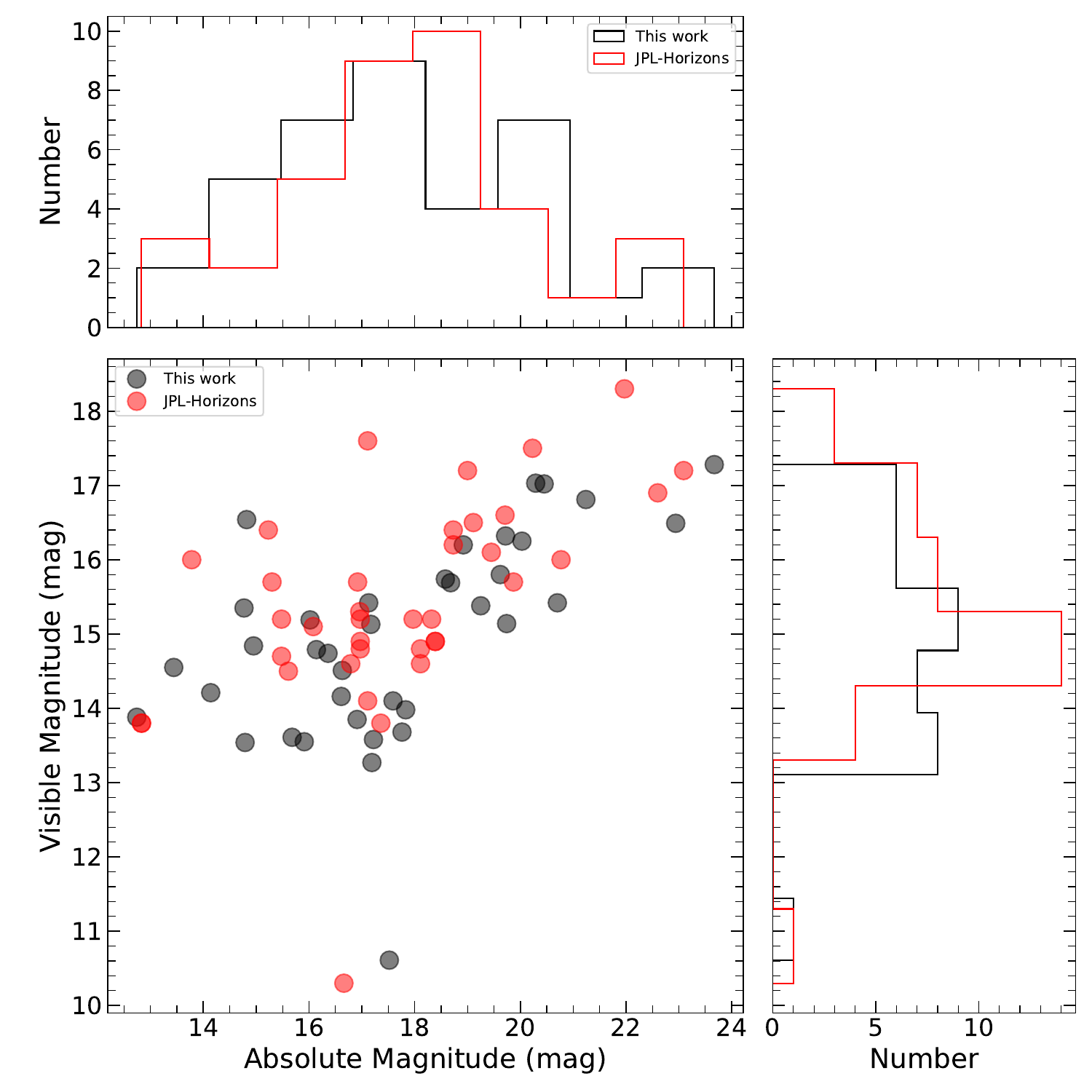}
    \caption{Visible (V-band) and absolute magnitudes ($H_{V}$) of the MIRSI+MOC NEO survey. In black circles are those magnitudes solutions derived in this work while red ones correspond to the JPL Horizons systems. The top and right sub-figures are the histogram distribution of the absolute and visible magnitudes, respectively.}
    \label{fig:HV}
\end{figure}


\begin{deluxetable*}{ccccccccccccc}[ht]
\tablenum{1}
\tablecaption{Summary of observed asteroids using MIRSI and MOC from the NASA-IRTF. \label{tab:Table1}}
\tablewidth{0pt}
\tablehead{
\colhead{Asteroid} & \colhead{Date} & \colhead{UT} & \colhead{Optical} & \colhead{IR} & \colhead{V} & \colhead{Airmass} & \colhead{r} & \colhead{$\Delta$} & \colhead{Phase}\\
\colhead{} & \colhead{} & \colhead{} & \colhead{STD} & \colhead{STD} & \colhead{[mag]} & \colhead{} & \colhead{[au]} & \colhead{[au]} & \colhead{[deg]}
}
\startdata
28P/Neujmin 1 & 2021-10-18 & 13:13 & - & Bad 16 & 17.03 & 1.1 & 2.89 & 2.25 & 17.4 \\
1994 PC1 & 2022-01-18 & 06:23 & SA94-702 & $\alpha$Tau & 10.31 & 1.3 & 0.99 & 0.02 & 79.9 \\
2002 AL14 & 2022-02-03 & 08:24 & BD+29 2091 & $\mu$UMa & 15.19 & 1.5 & 1.13 & 0.16 & 20.2 \\
2004 TY16 & 2022-02-03 & 09:39 & Wolf 365 & $\mu$UMa & 15.26 & 1.4 & 1.18 & 0.22 & 25.9 \\
Eger (1982 BB) & 2022-02-03 & 10:50 & Wolf 365 & $\mu$UMa & 15.7 & 1.1 & 1.49 & 0.54 & 17.6 \\
2006 DP14 & 2022-02-22 & 05:31 & Ross 889 & $\alpha$Tau & 17.23 & 2.7 & 1.23 & 0.25 & 12.9 \\
2004 TY16 & 2022-02-22 & 07:27 & G 163 50 & $\alpha$Hya & 14.82 & 2.5 & 1.18 & 0.2 & 17.5 \\
2001 SN263 & 2022-03-06 & 08:30 & LHS 0033 & $\beta$Gem & 14.12 & 1.1 & 1.06 & 0.1 & 47.5 \\
2004 TY16 & 2022-03-06 & 09:17 & G 163 51 & $\beta$Gem & 14.88 & 1.3 & 1.2 & 0.22 & 14.9 \\
1991 VK & 2022-03-06 & 10:20 & PG 1528+062B & $\beta$Gem & 15.33 & 2.5 & 1.12 & 0.19 & 44.6 \\
2002 DJ5 & 2022-03-27 & 05:42 & BD-12 2918 & $\mu$UMa & 15.74 & 1.5 & 1.05 & 0.07 & 41.2 \\
2000 EE14 & 2022-03-27 & 06:58 & LHS 1858 & $\mu$UMa & 17.57 & 1.4 & 1.01 & 0.35 & 78.1 \\
1989 ML & 2022-06-06 & 12:39 & SA 108 475 & $\alpha$Her & 16.1 & 1.5 & 1.12 & 0.12 & 22.6 \\
2000 NM & 2022-07-10 & 10:13 & SA 107 351 & $\alpha$Her & 14.75 & 2.1 & 1.19 & 0.27 & 44.0 \\
1996 AS1 & 2022-07-10 & 11:07 & Hilt 733 & $\alpha$Her & 14.9 & 2.3 & 1.12 & 0.11 & 20.6 \\
2000 NM & 2022-07-28 & 05:25 & G 15-24 & $\alpha$Her & 15.24 & 1.0 & 1.05 & 0.26 & 74.6 \\
1996 AS1 & 2022-07-28 & 07:10 & SA 111 1925 & $\alpha$Lyr & 14.91 & 1.4 & 1.12 & 0.11 & 18.7 \\
2001 CP44 & 2022-08-04 & 07:43 & Ross 711 & $\alpha$Her & 15.91 & 2.2 & 1.33 & 0.9 & 49.7 \\
Didymos (1996 GT) & 2022-09-27 & 10:34 & HD 28995 & $\beta$Cet & 14.57 & 2.5 & 1.04 & 0.08 & 54.0 \\
Sigurd (1992 CC1) & 2022-10-02 & 05:01 & SA 107 351 & $\eta$Sgr & 16.4 & 2.4 & 0.9 & 0.43 & 90.3 \\
Didymos (1996 GT) & 2022-10-06 & 13:56 & BD-21 0910 & $\beta$And & 14.8 & 1.4 & 1.03 & 0.07 & 67.0 \\
2001 CC21 & 2023-01-22 & 14:14 & GCRV 5951 & $\mu$UMa & 16.39 & 1.3 & 1.15 & 0.18 & 21.4 \\
2001 CC21 & 2023-02-07 & 14:00 & BD+54 1216 & $\alpha$Boo & 16.23 & 2.0 & 1.1 & 0.14 & 35.5 \\
Mithra (1987 SB) & 2023-03-19 & 05:51 & GCRV 5757 & $\beta$Gem & 14.53 & 1.1 & 1.14 & 0.23 & 47.7 \\
2000 FL10 & 2023-03-19 & 06:37 & BD-12 2918 & $\alpha$CMa & 15.72 & 1.5 & 1.16 & 0.23 & 41.1 \\
2005 TF49 & 2023-04-16 & 06:03 & BD+29 2091 & $\beta$Gem & 16.54 & 1.1 & 1.06 & 0.11 & 58.0 \\
Ivar (1929 SH) & 2023-04-16 & 07:43 & Feige 66 & $\mu$UMa & 13.75 & 1.1 & 1.72 & 0.76 & 14.4 \\
2023 GM & 2023-04-16 & 08:21 & BD+9 2190 & $\mu$UMa & 16.93 & 1.3 & 1.01 & 0.02 & 67.8 \\
Ivar (1929 SH) & 2023-04-23 & 11:27 & Ross 838 & $\alpha$Her & 13.76 & 1.2 & 1.68 & 0.75 & 19.0 \\
Apollo (1932 HA) & 2023-04-23 & 12:00 & Ross 838 & $\alpha$Boo & 15.13 & 1.5 & 1.31 & 0.32 & 16.2 \\
Zeus (1988 VP4) & 2023-06-07 & 06:02 & - & $\mu$UMa & 16.74 & 1.3 & 1.35 & 0.6 & 44.4 \\
1994 XD & 2023-06-07 & 07:30 & - & $\mu$UMa & 14.57 & 1.7 & 1.07 & 0.06 & 25.5 \\
2005 EK70 & 2024-01-29 & 13:52 & - & $\mu$UMa & 16.27 & 1.2 & 1.09 & 0.21 & 55.9 \\
2006 AT2 & 2024-01-29 & 14:43 & - & $\mu$Uma & 15.95 & 1.5 & 1.23 & 0.28 & 25.6 \\
23 Thalia & 2024-01-29 & 15:23 & - & $\mu$UMa & 10.36 & 1.1 & 2.07 & 1.3 & 21.6 \\
2001 CC21 & 2024-02-02 & 07:45 & - & $\alpha$Tau & 18.56 & 1.1 & 1.17 & 0.34 & 49.5 \\
1992 BF & 2024-02-07 & 12:01 & GCRV 5951 & $\mu$UMa & 16.73 & 1.3 & 1.12 & 0.14 & 17.9 \\
2019 CC5 & 2024-02-07 & 13:51 & BD+26 2606 & $\mu$UMa & 18.33 & 1.1 & 1.0 & 0.06 & 72.2 \\
2023 SP1 & 2024-02-14 & 09:42 & BD+25 1981 & $\beta$Gem & 16.03 & 1.1 & 1.04 & 0.06 & 24.2 \\
2024 CG2 & 2024-02-14 & 11:28 & SA 103 626 & $\mu$UMa & 17.24 & 1.2 & 1.02 & 0.03 & 35.0 \\
2005 EK70 & 2024-02-14 & 12:23 & SA 103 626 & $\mu$UMa & 13.83 & 1.0 & 1.09 & 0.11 & 22.4 \\
1999 KV4 & 2024-03-20 & 15:07 & - & $\gamma$Dra & 15.82 & 1.1 & 1.06 & 0.21 & 67.4 \\
\enddata
\end{deluxetable*}

\section{First Results from MIRSI NEO Survey}

\begin{figure}[!]
    \centering
    \includegraphics[width=0.47\textwidth]{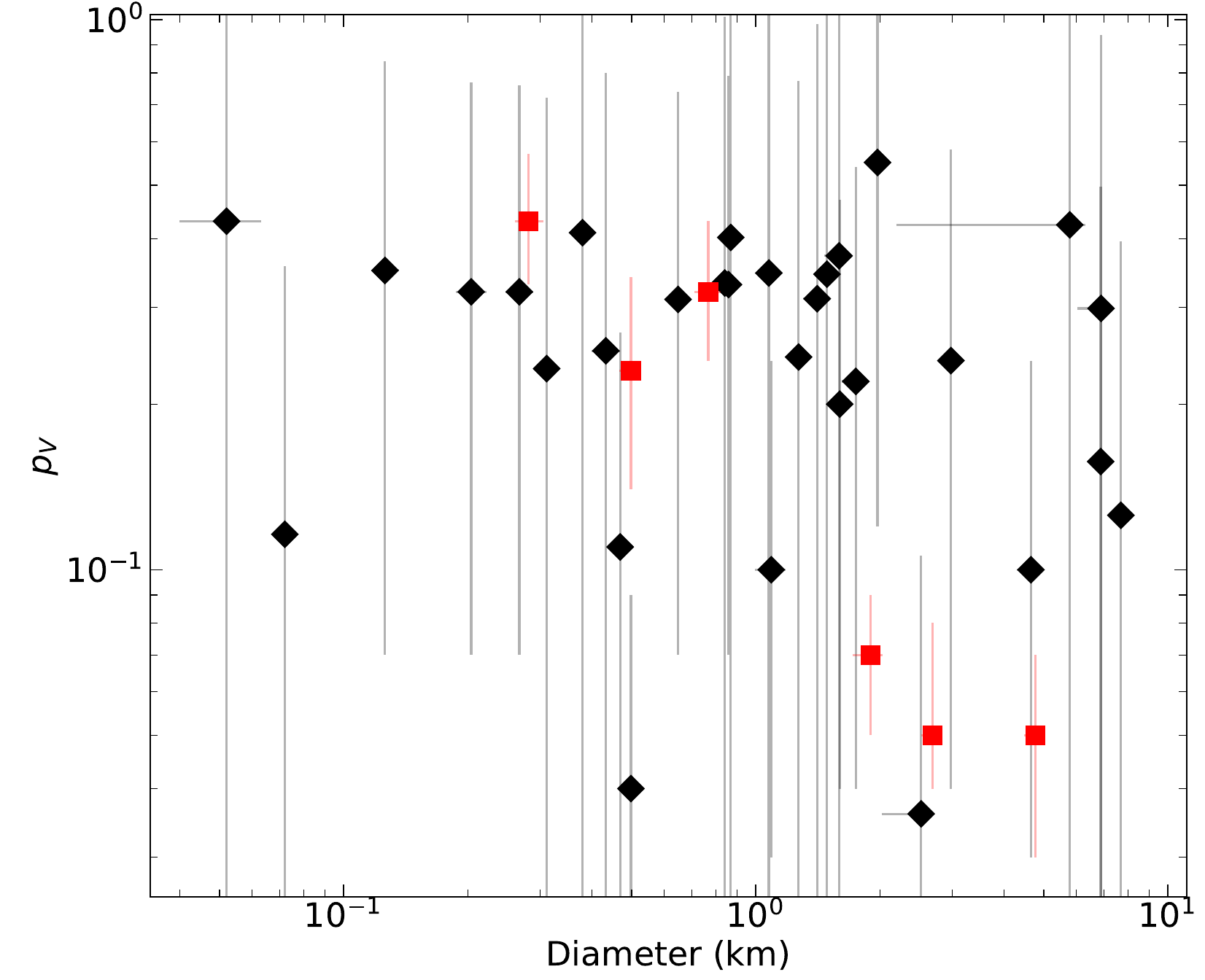}
    \caption{Visible albedo as a function of diameter for the 31 NEOs observed with MIRSI. The dark diamonds show the results using the MOC-$H_{V}$ while the red ones show the solutions with JPL-$H_{V}$.}
    \label{fig:D_pv}
\end{figure}

In this paper, we report the results of the first 31 NEOs and 2 main-belt asteroid observations from our simultaneous optical and thermal characterization campaign with MIRSI. We performed 44 observations of 33 unique targets observed from 2022 to 2024 (see Table~\ref{tab:Table1} for observational circumstances). As shown in Figure \ref{fig:HV}, most of these asteroids' visible magnitudes ranged between 14 and 16~mags during their observations. The majority of the observations were carried out while MIRSI had installed the engineering IR detector, which was replaced in early 2024. We were able to obtain simultaneous optical and thermal photometry for 25 of the 31 NEOs observed in this program. Our sample contains objects that range from 15~mJy to hundreds of Jansky with a median of 330~mJy. In this section, we primarily describe results from the N-band analysis, whereas in \S \ref{sec:indiv_targets}, we discuss in detail the results and implications of targets of interest, such as mission support observations and future MIRSI applications for NEO studies.

\begin{figure}[h!]
    \centering
    \includegraphics[width=0.47\textwidth]{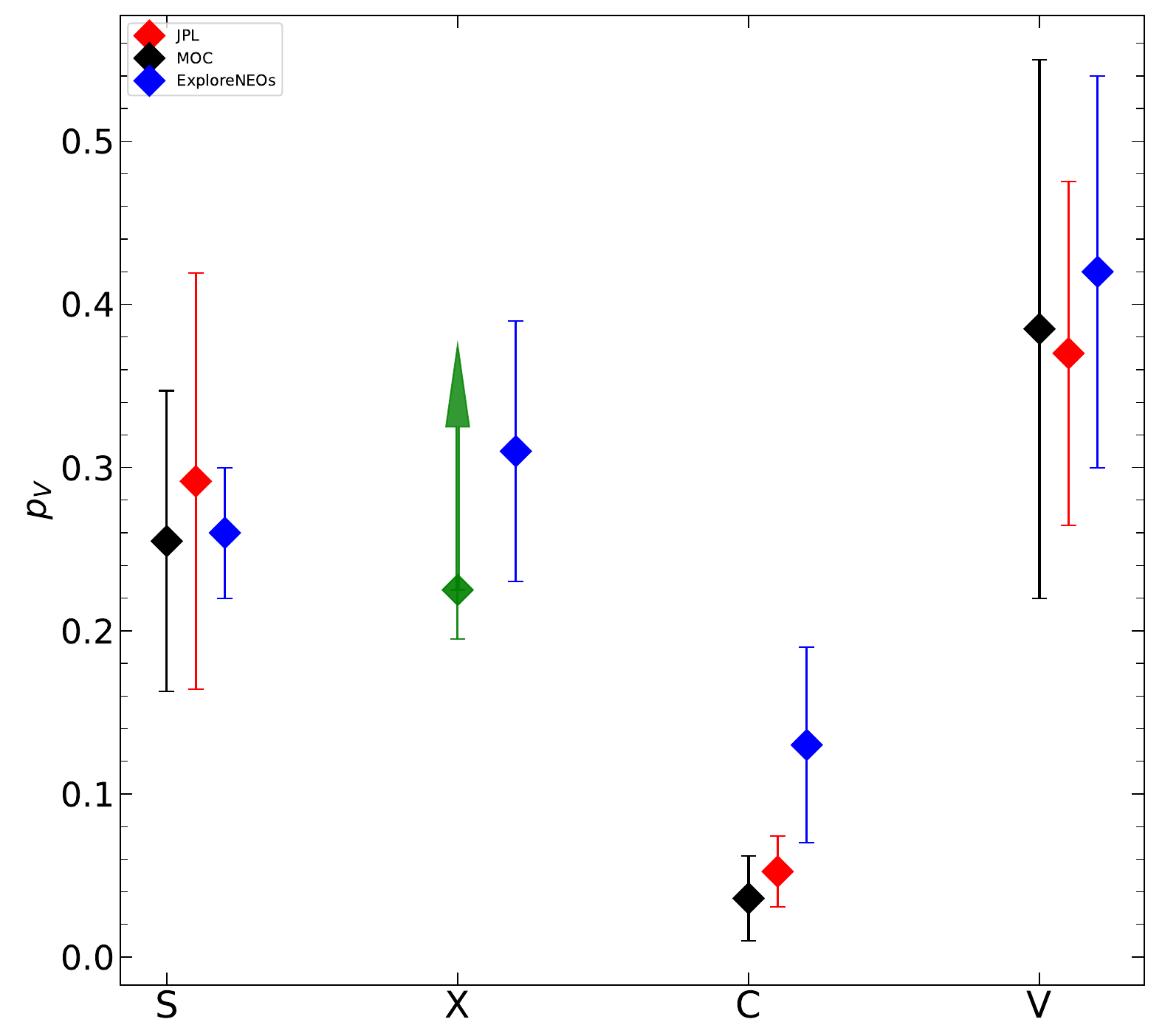}
    \caption{Average visible albedo of NEOs by taxonomic type. Black and red diamonds illustrate the average $p_{V}$ using $H_{V}$ from MOC and JPL, respectively. In blue diamonds, we show the comparable average albedos for complex taxonomies from previous ExploreNEOs work by \citet{Thomas2011}. In green, we show the average and the range of albedos for X-complex based on our estimates from weakly detected objects.}
    \label{fig:Taxo_pv}
\end{figure}

Table \ref{tab:Table2} shows the main results of this commissioning science campaign, including the MOC measured absolute magnitude ($H_{V}$) and diameter ($D$) and albedo ($p_{V}$) solutions using both MOC and JPL $H_{V}$ (see Figure \ref{fig:HV}). Those results listed in Table \ref{tab:Table2} correspond to the single N-band analysis, and the uncertainties in $D$ and $p_{V}$ correspond to the 1$\sigma$ values. As illustrated in Figure \ref{fig:D_pv}, our sample contains objects ranging from tens of meters to a few kilometers in size and derived albedos from 4\% to 53\%. We measured the diameter and albedo for 11 NEOs whose properties were unknown, including three recently discovered NEOs.

Since about 85\% of the objects included in our survey had previous taxonomic classification, we compared our measured albedos with those obtained by the ExploreNEOs team \citep{Thomas2011}. In Figure \ref{fig:Taxo_pv}, we show the average $p_{V}$ for S-, C-, V-, and X-types taxonomies obtained from 19, 4, 1, and 2 objects, respectively. $p_{V}$ solutions using both JPL and MOC $H_{V}$ yield relatively similar averaged values for taxonomic types, and all of the averaged solutions agree within error with those reported by \citet{Thomas2011}. We note that our V-type average is the weighted average from two different measurements of the same NEO 2000 NM. 

For six NEOs, we only provide albedo and diameter using the JPL $H_{V}$ as we could not collect MOC photometry for them. Since this is the first MIRSI-NEO program running simultaneous optical and thermal photometry, we decided to provide $p_{V}$ solutions using both MOC and JPL $H_{V}$ particularly since, in cases, our optical photometry was slightly compromised by weather or the targets were too faint. Thus, for those objects, identified as having poor optical photometry in Table \ref{tab:Table2}, the $p_{V}$ solution using the $H_{V}$ from JPL is highly favored. About half of the objects with MOC photometry showed to differ by or more than 0.3~mags from the JPL-H solution. On average, we measured a lower $H_{V}$ than those reported by JPL-Horizons by 0.06~mags.

\section{Discussion \label{sec:indiv_targets}}

\subsection{Diameters and albedos determined with the MIRSI system} \label{sec:diam_pv_disc}

In Figure \ref{fig:D_pv} we illustrate the diameters and albedos of the observed NEOs with MIRSI. The first outcome of visualizing such figure is the presence of objects with $p_{V}$ near or larger than 50\%. \citet{Gustafsson_2019} demonstrated that objects with $p_{V}>$50\% are likely indicative of inaccuracies in the measured absolute magnitude or diameter rather than an actual physical property of the object. While errors in absolute magnitude are the primary source of albedo uncertainties \citep{Masiero2021}, a wrongly measured diameter can also affect the albedo determination. For example, underestimating the diameter leads to an overestimation of albedo. 21 of the 31 unique NEOs observed in our program have previous diameter measurements (see Table \ref{tab:Table2}). After excluding those measurements with poor detections, we find an average ratio of $D(us)/D(literature)=$ 0.95 by comparing our diameter estimates divided by those published values. When dividing our albedo measurements by those in the literature, we obtain an average ratio of $p_{V}(us)/p_{V}(literature)=$1.3. From this small comparison with 21 objects, the diameters measured with the MIRSI system seem to be underestimated, indicating a tendency toward albedo overestimation. These results are broadly consistent with the findings of \citet{Harris_2011}, who reported that diameters and albedos derived from single-band thermal observations and the NEATM yield measurements accurate to within 20\% and 50\%, respectively, when compared to literature values.  

There are a few factors that could be leading to the underestimation of the diameter using MIRSI. The flux calibration could be a dominant factor, especially for faint targets where longer observations are needed in order to be detected, which is the case for many of our objects. We have seen typical flux underestimations of about 5\% in our calibrations due to sky variability. Ultimately, this can result in an underestimation of the diameter of as much as 25\%. An additional factor contributing to the diameter underestimation comes from the assumption of the beaming parameter (see \S \ref{sec:TherMod}). We note that on occasions, our thermal modeling did not provide a good fit to the observed data, such as in the case of NEO 2005 TY16. We found that higher beaming was preferred in order to yield a diameter that resulted in an acceptable albedo ($<5$\%) using the object's taxonomy (D-type) as a prior.

We point out that high $p_{V}$ solutions in our survey could be induced by wrongly determined $H_{V}$. Our $H_{V}$ estimation is based on a single observation with limited coverage of the asteroid phase curve and assumed photometric variables (i.e., the $G$ slope parameter). Therefore, MOC photometry is strongly preferred for asteroids with limited photometric data, particularly those that have been recently discovered. Asteroids with $H_{V}$ values robustly determined from a compilation of multiple photometric observations are strongly preferred \citep{MAHLKE2021114094}.

\subsection{Discussion on individual targets}

\subsubsection{7482 (1994 PC1): Simultaneous Optical and Thermal lightcurves}
\begin{figure}[t]
    \centering
    \includegraphics[width=0.47\textwidth]{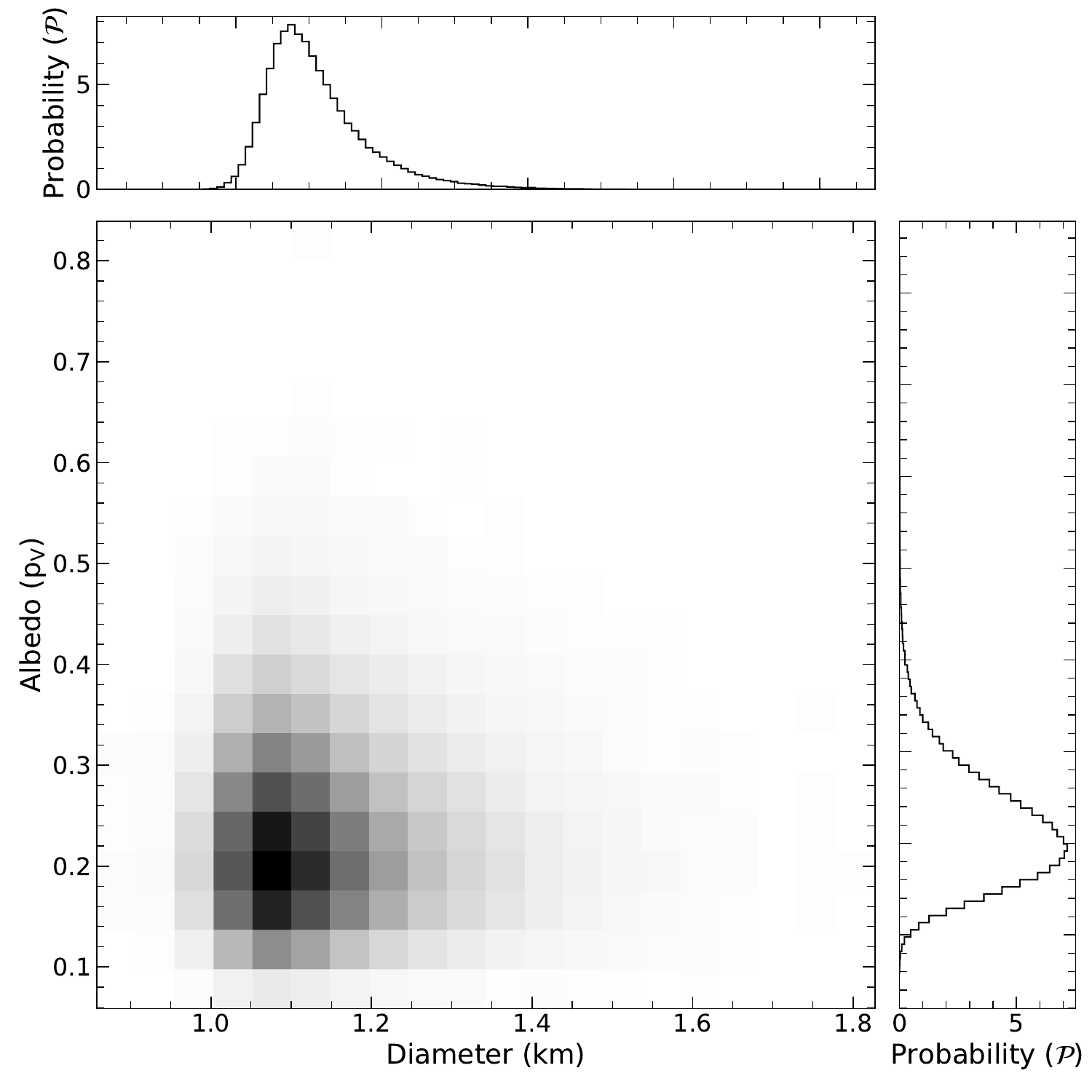}
    \caption{Two-dimensional histogram of the posterior probability function for albedos and diameters solutions for the NEO 7482. Darker shades indicate more probable values. Top and right sub-panels are the one-dimensional marginal distributions of the likelihood probability function for the diameter and albedo, respectively.}
    \label{fig:2D_hist}
\end{figure}

The asteroid 7482 was the first NEO observed by the MIRSI system on 2022 January 18, when it was 0.02~AU from Earth and with a brightness of 10.3~mags at V-band. For such geometry, we predicted the thermal flux of 7482 to be around 60~Jy in the N-band. We observed 7482 with MIRSI's 7.7, 8.7, 11.7, and 12.5~${\mu}$m filters and measured fluxes of 59.8$\pm$1.5, 64.1$\pm$0.8, 65.2$\pm$0.6, and 73.6$\pm$1.3~Jy, respectively. We obtained best-fit thermal model parameters of $D=$1.091$^{+0.077}_{-0.046}$, $p_V=$0.21$^{+0.07}_{-0.05}$, and $\eta=$0.72$^{+0.12}_{-0.06}$ (see Figure \ref{fig:2D_hist}). These values agree with the 0.28$\pm$0.19 albedo and 1.052$\pm$0.30~km diameter from NEOWISE \citep{Nugent_2015}. In addition, our 21\% albedo matches the S-type taxonomy of 7482 \citep{Binzel2019}. 

\begin{figure}[h!]
    \centering
    \includegraphics[width=0.47\textwidth]{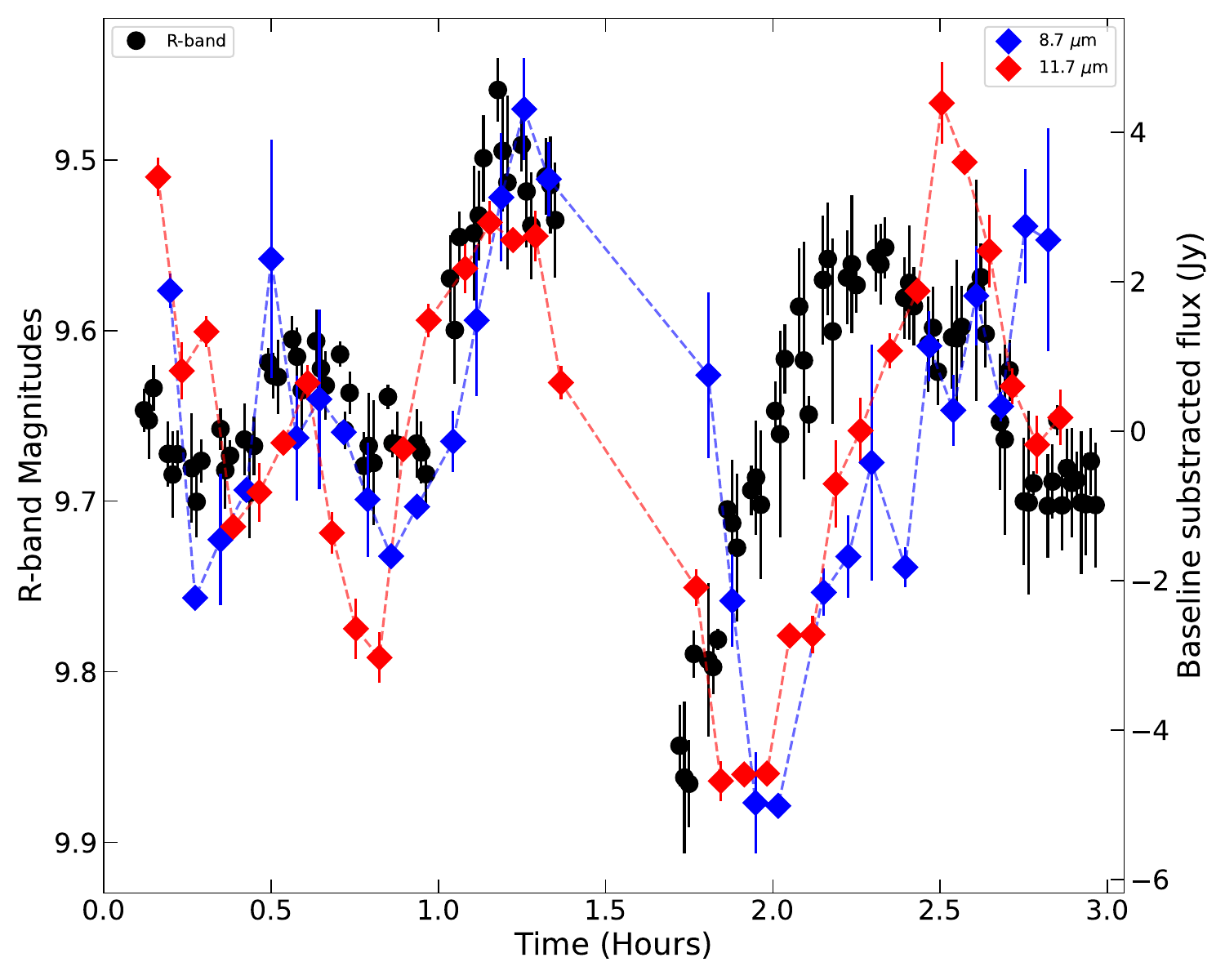}
    \caption{Simultaneous optical and thermal lightcurves of NEO 7482 as a function of time (hours) after the beginning of the observation on 2022 January 18. The left y-axis illustrates the R-band magnitudes (black circles). The right y-axis shows the flux variability at the 8.7~$\mu$m (blue diamonds) and 11.7~$\mu$m (red diamonds) filters.}
    \label{fig:7482_lc}
\end{figure}

As discussed in \S \ref{sec:observation}, MIRSI's system allows for simultaneous optical and thermal observations. Because 7482 was bright enough to be detected within a single MIRSI exposure at different filters, we were able to obtain thermal lightcurves at 8.7 and 11.7~${\mu}$m and R-band photometry across the entire 2.599~h rotation period of the object \citep{Franco2022}. In Figure \ref{fig:7482_lc}, we show the simultaneous optical R-band and thermal lightcurves of 7482. From the optical lightcurve amplitude, we obtained an a/b ratio of 1.2 using \citet{Kwiatkowski2010} approach, suggesting a somewhat aspherical object. Analyzing the thermal flux variability of the 8.7 and 11.7~${\mu}$m filters suggests an effective diameter variability of around 100~meters from peak (maxima) to peak (minima).



\begin{longrotatetable}
\begin{deluxetable*}{ccccccccccccc}
\tablenum{2}
\tablecaption{Results of our MIRSI and MOC N-band survey from the NASA IRTF. \label{tab:Table2}}
\tablewidth{0pt}
\tablehead{
\colhead{Asteroid} & \colhead{H$_{MOC}$} & \colhead{H$_{JPL}$} & \colhead{Diff.} & \colhead{Flux$_{10.5}$} & \colhead{$\eta$} & \colhead{Tax.} & \multicolumn{2}{c}{MOC} & \multicolumn{2}{c}{JPL} & \multicolumn{2}{c}{Literature}\\ 
\cmidrule(lr){8-9}\cmidrule(lr){10-11}\cmidrule(lr){12-13}
\colhead{} & \colhead{} & \colhead{} & \colhead{} & \colhead{} & \colhead{} & \colhead{} & \colhead{D} & \colhead{p$_{V}$} & \colhead{D} & \colhead{p$_{V}$} & \colhead{D} & \colhead{p$_{V}$}\\
\colhead{} & \colhead{(mag)} & \colhead{(mag)} & \colhead{(mag)} & \colhead{(mJy)} & \colhead{} & \colhead{} & \colhead{(km)} & \colhead{} & \colhead{(km)} & \colhead{} & \colhead{(km)} & \colhead{}
}
\startdata
28 (Neujmin 1) & - & 10.3 & - & 197$\pm$14 & 0.93 & S$^{f}$ & - & - & 26.29$_{-5.421}^{+4.936}$ & 0.19$_{-0.05}^{+0.06}$ & 34.56$\pm$0.72 & 0.13$\pm$0.01 \\
506459 (2002 AL14) & 16.91$\pm$0.14 & 17.97 & 1.06 & 212$\pm$20 & 0.96 & S/L & 0.870$_{-0.039}^{+0.044}$ & 0.40$_{-0.49}^{+0.86}$ & 0.878$_{-0.043}^{+0.046}$ & 0.28$_{-0.11}^{+0.15}$ & 0.785$\pm$0.188$^{d}$ & 0.18$\pm$0.1$^{d}$ \\
170891 (2004 TY16) & 15.68$\pm$0.13 & 16.97 & 1.29 & 430$\pm$20 & 1.01 & D & 1.593$_{-0.125}^{+0.101}$ & 0.37$_{-0.46}^{+0.8}$ & 1.583$_{-0.067}^{+0.056}$ & 0.18$_{-0.07}^{+0.09}$ & 3.346$^{d}$ & 0.03$^{j}$ \\
3103 (Eger (1982 BB) & 14.14$\pm$0.13$^{a}$ & 15.3 & 1.16 & 10$\pm$0$^{*}$  & 0.93 & Xe & $<$2.1 & $>$0.2 & $<$2.1 & $>$0.2 & 1.576$^{b}$ & 0.42$^{b}$ \\
388188 (2006 DP14) & 17.17$\pm$0.74 & 19.0 & 1.83 & 30$\pm$0$^{*}$ & 0.89 & S$^{h}$ & $<$0.65 & $>$0.25 & $<$0.65 & $>$0.25 & 0.4 & - \\
153591 (2001 SN263) & 17.22$\pm$0.22 & 17.11 & -0.11 & 6872$\pm$18 & 1.22 & C/B & 2.516$_{-0.5}^{+0.0}$ & 0.03$_{-0.04}^{+0.07}$ & 2.523$_{-0.182}^{+0.103}$ & 0.04$_{-0.01}^{+0.02}$ & 2.127$^{b}$ & 0.05$^{b}$ \\
170891 (2004 TY16) & 15.91$\pm$0.08 & 16.97 & 1.06 & 404$\pm$14 & 0.9 & D & 1.489$_{-0.098}^{+0.095}$ & 0.34$_{-0.42}^{+0.75}$ & 1.116$_{-0.065}^{+0.063}$ & 0.23$_{-0.09}^{+0.11}$ & 3.346$^{d}$ & 0.03$^{d}$ \\
7341 (1991 VK) & 16.63$\pm$0.15 & 16.96 & 0.33 & 267$\pm$17 & 1.19 & Q & 1.271$_{-0.063}^{+0.066}$ & 0.24$_{-0.3}^{+0.53}$ & 1.105$_{-0.06}^{+0.068}$ & 0.24$_{-0.09}^{+0.12}$ & 0.982$\pm$0.316 & 0.23$\pm$0.11 \\
317255 (2002 DJ5) & 19.74$\pm$0.29 & 19.87 & 0.13 & 178$\pm$16 & 1.16 & S & 0.311$_{-0.021}^{+0.018}$ & 0.23$_{-0.28}^{+0.49}$ & 0.281$_{-0.018}^{+0.021}$ & 0.25$_{-0.09}^{+0.13}$ & 0.276$^{b}$ & 0.2$^{b}$ \\
138127 (2000 EE14) & -$^{a}$  & 17.11 & - & 12$\pm$0$^{*}$ & 1.51 & Sq & $<$1.5 & $>$0.3 & $<$1.5 & $>$0.3 & 0.754$\pm$0.023 & 0.45$\pm$0.06 \\
10302 (1989 ML) & 20.03$\pm$0.16 & 19.45 & -0.58 & 22$\pm$0$^{*}$ & 0.98 & X & $<$0.5 & $>$0.25 & $<$0.5 & $>$0.25 & 0.203$^{b}$ & 0.56$^{b}$ \\
215188 (2000 NM) & 14.79$\pm$0.09 & 15.48 & 0.69 & 506$\pm$14 & 1.19 & V$^{g}$ & 1.975$_{-0.131}^{+0.151}$ & 0.55$_{-0.43}^{+0.76}$ & 1.699$_{-0.152}^{+0.133}$ & 0.38$_{-0.09}^{+0.12}$ & - & - \\
23606 (1996 AS1) & 17.83$\pm$0.12 & 18.39 & 0.56 & 755$\pm$14 & 0.96 & S & 0.648$_{-0.047}^{+0.049}$ & 0.31$_{-0.24}^{+0.43}$ & 0.618$_{-0.041}^{+0.037}$ & 0.2$_{-0.05}^{+0.06}$ & 0.866$\pm$0.013 & 0.12$\pm$0.02 \\
215188 (2000 NM) & 16.02$\pm$0.82 & 15.48 & -0.54 & 331$\pm$14 & 1.48 & V & 1.75$_{-0.13}^{+0.117}$ & 0.22$_{-0.18}^{+0.32}$ & 1.763$_{-0.118}^{+0.119}$ & 0.36$_{-0.09}^{+0.12}$ & - & - \\
23606 (1996 AS1) & 19.62$\pm$0.33 & 18.39 & -1.23 & 436$\pm$14 & 0.94 & S & 0.469$_{-0.029}^{+0.029}$ & 0.11$_{-0.09}^{+0.16}$ & 0.493$_{-0.034}^{+0.037}$ & 0.31$_{-0.08}^{+0.1}$ & 0.866$\pm$0.013 & 0.2 \\
25916 (2001 CP44) & 14.82$\pm$0.15 & 13.78 & -1.04 & 280$\pm$20 & 1.24 & Sw & 4.655$_{-0.35}^{+0.303}$ & 0.1$_{-0.07}^{+0.14}$ & 4.658$_{-0.363}^{+0.382}$ & 0.27$_{-0.07}^{+0.09}$ & 5.683$\pm$0.03 & 0.26$\pm$0.05 \\
11066 (Sigurd (1992 CC1)) & 14.77$\pm$0.15 & 15.23 & 0.46 & 251$\pm$50 & 1.63 & S & 2.976$_{-0.209}^{+0.227}$ & 0.24$_{-0.2}^{+0.34}$ & 2.761$_{-0.195}^{+0.179}$ & 0.18$_{-0.05}^{+0.06}$ & 2.104$\pm$0.085 & 0.33$\pm$0.06 \\
98943 (2001 CC21) & 18.92$\pm$0.12$^{a}$  & 18.73 & -0.19 & 73$\pm$10 & 0.97 & S$^{i}$ & 0.433$_{-0.034}^{+0.036}$ & 0.25$_{-0.32}^{+0.55}$ & 0.437$_{-0.027}^{+0.03}$ & 0.30$_{-0.12}^{+0.15}$ & 0.465$\pm$0.015$^{e}$ & 0.22$\pm$0.02$^{e}$ \\
98943 (2001 CC21) & 18.68$\pm$0.24 & 18.73 & 0.05 & 53$\pm$15$^{*}$ & 1.1 & S$^{i}$ & 0.304$_{-0.023}^{+0.024}$ & 0.64$_{-0.5}^{+0.89}$ & 0.291$_{-0.025}^{+0.026}$ & 0.64$_{-0.16}^{+0.21}$ & 0.465$\pm$0.015$^{e}$ & 0.22$\pm$0.02$^{e}$ \\
4486 (Mithra (1987 SB)) & 16.36$\pm$0.18$^{a}$ & 15.61 & -0.75 & 684$\pm$32 & 1.22 & Sq & 1.599$_{-0.113}^{+0.098}$ & 0.2$_{-0.16}^{+0.27}$ & 1.656$_{-0.141}^{+0.14}$ & 0.35$_{-0.09}^{+0.11}$ & 1.849$\pm$0.022 & 0.3$\pm$0.06 \\
86666 (2000 FL10) & 17.13$\pm$0.22 & 16.92 & -0.21 & 150$\pm$32 & 1.16 & S & 0.859$_{-0.067}^{+0.066}$ & 0.33$_{-0.26}^{+0.46}$ & 0.819$_{-0.068}^{+0.074}$ & 0.43$_{-0.11}^{+0.14}$ & 1.165$\pm$0.456 & 0.23$\pm$0.13 \\
2005TF49 (2005 TF49) & 20.29$\pm$0.19 & 19.11 & -1.18 & 30$\pm$10$^{*}$ & 1.32 & - & 0.204$_{-0.017}^{+0.018}$ & 0.32$_{-0.25}^{+0.45}$ & 0.22$_{-0.015}^{+0.017}$ & 0.82$_{-0.2}^{+0.26}$ & - & - \\
1627 (Ivar (1929 SH)) & 13.44$\pm$0.14 & 12.83 & -0.61 & 560$\pm$22 & 0.9 & S & 7.693$_{-0.449}^{+0.413}$ & 0.13$_{-0.16}^{+0.27}$ & 7.62$_{-0.865}^{+0.546}$ & 0.22$_{-0.13}^{+0.17}$ & 8.37$\pm$0.075 & 0.13$\pm$0.03 \\
2023GM (2023 GM) & 23.67$\pm$0.11 & 22.6 & -1.07 & 63$\pm$10$^{*}$ & 1.41 & - & 0.072$_{-0.005}^{+0.005}$ & 0.12$_{-0.05}^{+0.24}$ & 0.073$_{-0.006}^{+0.006}$ & 0.11$_{-0.06}^{+0.07}$ & - & - \\
1627 (Ivar (1929 SH)) & 12.74$\pm$0.18 & 12.83 & 0.09 & 657$\pm$30 & 0.94 & S & 6.886$_{-3.59}^{+0.515}$ & 0.29$_{-0.53}^{+0.91}$ & 6.915$_{-1.491}^{+3.973}$ & 0.27$_{-0.14}^{+0.19}$ & 8.37$\pm$0.075 & 0.13$\pm$0.03 \\
1862 (Apollo (1932 HA)) & 16.14$\pm$0.15 & 16.08 & -0.06 & 224$\pm$15 & 0.92 & Q & 1.410$_{-0.105}^{+0.096}$ & 0.31$_{-0.39}^{+0.67}$ & 1.387$_{-0.08}^{+0.078}$ & 0.33$_{-0.13}^{+0.17}$ & 1.264$\pm$0.037 & 0.35$\pm$0.04 \\
5731 (Zeus (1988 VP4)) & - & 15.48 & - & 827$\pm$32 & 1.19 & C & - & - & 4.773$_{-0.286}^{+0.218}$ & 0.05$_{-0.02}^{+0.02}$ & 5.231$\pm$0.686 & 0.03$\pm$0.01 \\
488453 (1994 XD) & - & 19.28 & - & 435$\pm$20 & 1.01 & S & - & - & 0.281$_{-0.021}^{+0.025}$ & 0.43$_{-0.1}^{+0.14}$ & - & - \\
187026 (2005 EK70) & - & 17.4 & - & 130$\pm$13 & 1.3 & Sq & - & - & 0.767$_{-0.056}^{+0.055}$ & 0.32$_{-0.08}^{+0.11}$ & - & - \\
417264 (2006 AT2) & - & 17.13 & - & 1110$\pm$22 & 1.01 & C & - & - & 1.899$_{-0.182}^{+0.13}$ & 0.07$_{-0.02}^{+0.02}$ & 1.94$\pm$0.014 & 0.06 \\
98943 (2001 CC21) & - & 18.74 & - & 15$\pm$0$^{*}$ & 1.24 & S$^{i}$ & - & - & 0.398$_{-0.031}^{+0.027}$ & 0.35$_{-0.09}^{+0.11}$ & 0.465$\pm$0.015$^{e}$ & 0.22$\pm$0.02$^{e}$ \\
152563 (1992 BF) & 19.72$\pm$0.14 & 19.71 & -0.01 & 86$\pm$8 & 0.93 & S & 0.267$_{-0.02}^{+0.022}$ & 0.32$_{-0.25}^{+0.44}$ & 0.261$_{-0.017}^{+0.02}$ & 0.3$_{-0.07}^{+0.1}$ & 0.272$\pm$0.077 & 0.29$\pm$0.19 \\
2019CC5 (2019 CC5) & 21.24$\pm$0.13 & 21.97 & 0.73 & 33$\pm$5$^{*}$ & 1.46 & - & 0.126$_{-0.008}^{+0.01}$ & 0.35$_{-0.28}^{+0.49}$ & 0.104$_{-0.008}^{+0.009}$ & 0.25$_{-0.06}^{+0.08}$ & - & - \\
2023SP1 (2023 SP1) & 20.7$\pm$0.24 & 20.77 & 0.07 & 567$\pm$17 & 0.99 & - & 0.498$_{-0.027}^{+0.025}$ & 0.04$_{-0.03}^{+0.05}$ & 0.494$_{-0.03}^{+0.026}$ & 0.03$_{-0.01}^{+0.01}$ & - & - \\
2024CG2 (2024 CG2) & 22.94$\pm$0.15 & 23.09 & 0.15 & 14$\pm$9$^{*}$ & 1.1 & - & 0.052$_{-0.012}^{+0.011}$ & 0.43$_{-0.33}^{+0.32}$ & 0.053$_{-0.013}^{+0.011}$ & 0.38$_{-0.13}^{+0.18}$ & - & - \\
187026 (2005 EK70) & 17.19$\pm$0.15 & 17.36 & 0.17 & 568$\pm$29 & 0.98 & Sq & 0.691$_{-0.06}^{+0.047}$ & 0.49$_{-0.39}^{+0.68}$ & 0.607$_{-0.054}^{+0.054}$ & 0.53$_{-0.13}^{+0.17}$ & - & - \\
25330 (1999 KV4) & - & 16.75 & - & 1967$\pm$30 & 1.41 & B & - & - & 2.684$_{-0.167}^{+0.115}$ & 0.05$_{-0.01}^{+0.03}$ & 2.7$\pm$0.75$^{c}$ & 0.08$\pm$0.02$^{c}$ \\
\hline
Ratio (us/literature) &  &  & & &  &  &  &  & &  & 0.95 & 1.29 \\
\enddata

\tablecomments{The column Diff. indicate the magnitude difference between JPL and MOC. Asteroid fluxes marked with ``*'' indicate poor detections (i.e., below 5$\sigma$). Absolute magnitudes from MOC ($H_{MOV}$) marked with ``a'' indicate nights with variable weather. The values under the literature column marked with ``b'' correspond to albedo and diameter measurements from the ExploreNEOs program \citep{Trilling_2010, Mueller_2011, Trilling_2016}, ``c'' from \citet{DELBO2003116}, ``d'' from the Solar system Open Database Network \citet{Berthier2023}, ``e'' from \citet{Fornasier2024}, ``j'' from \citet{Binzel2019}, and those without a mark from \citet{Mainzer2011}. Object taxonomies marked with ``f'' were obtained from \citet{Tholen1989}, ``g'' from \citet{Licandro2008}, ``h'' from \citet{Somers2010}, ``i'' from \citet{Geem2023} and those without a mark from \citet{Binzel2019}.}
\end{deluxetable*}
\end{longrotatetable}

\citet{Morrison1976} demonstrated that thermal inertia can be measured by analyzing phase lags between simultaneously obtained thermal and optical lightcurves. A thermal light curve will shift from the optical if the asteroid regolith has non-zero thermal inertia. However, several parameters can influence phase lags and not necessarily the asteroid's thermal properties. For example, complex-shaped objects with craters or bifurcated structures. Thus, since the shape of 7482 is unknown, we cannot argue whether the phase lag observed around two hours in Figure \ref{fig:7482_lc} is dominated by complexities in shape or non-zero thermal inertia. We leave extensive thermophysical modeling analysis of 7482 for future work once shape models become available from the 2022 close approach. We insist that this capability of the MIRSI system could be a powerful technique for regolith characterization of NEOs with shape models.

\subsubsection{Mission Support Observations: Characterization of 98943 (2001 CC21)}
In 2026, the Japan Aerospace Exploration Agency (JAXA) Extended Hayabusa2 Mission (named Hayabusa2$\sharp$) will fly by the NEO (98943) Torifune 2001~CC21 \citep[hereafter 98943;][]{MIMASU2022557}. Physical properties such as diameter and albedo are crucial for the safe and successful planning of the fly-by. For example, knowing 98943's reflectivity properties, such as the albedo, can better inform JAXA about what adjustments would be needed on instrument calibrations (i.e., needed integration times to achieve a desired S/N) to achieve a better outcome during the fly-by, especially since Hayabusa2 was planned for the B-type asteroid Ryugu, which is known to hold a much darker surface than 98943. The asteroid 98943 was initially taxonomic classified as an L-type \citep{Binzel2004}, but subsequent spectroscopy \citep{Demeo2009, Lazzarin2005} and polarimetry \citep{Geem2023} characterizations support an S-type due to the apparent absorption features at 0.9 and 1.9~${\mu}$m.

In this program, we performed three observations of 98943 to better constrain its diameter and $p_{V}$. Our optical and thermal observations obtained on 2023 January 22 yielded an $H_{V}$ estimate of 18.93$\pm$0.12 and an N-band flux of 73$\pm$10~mJy. Using our measured $H_{V}$, we obtained best-fit diameter and $p_{V}$ of 0.433$_{-0.034}^{+0.036}$~km and 0.25$_{-0.32}^{+0.55}$, respectively. Using the JPL-Horizons $H_{V}$, we obtained best-fit $p_{V}$ of 0.30$_{-0.12}^{+0.15}$. Additional follow-up observations were made on 2023 February 7 and 2024 February 2. On 2023 February 7, we obtained best-fit diameter and $p_{V}$ of 0.304$_{-0.023}^{+0.024}$~km and 0.64$_{-0.16}^{+0.21}$, while on 2024 February 2 we obtained best-fit diameter and $p_{V}$ of 0.398$_{-0.029}^{+0.025}$~km and 0.35$_{-0.09}^{+0.11}$. However, since the object was fainter and the sky conditions were highly variable, those observations resulted in weaker detections (i.e., below 3$\sigma$). Therefore, we interpret these measurements with caution and exclude them from our discussion.

Our best solution of $D=433$~meters is about half the effective diameter that was initially assumed for 98943 \citep{HIRABAYASHI20211533}. Recent characterization campaigns by \citep{Fornasier2024} and \citet{Popescu2025} seem to agree with a diameter ranging from 0.43 to 0.48~km. We clarify that our diameter estimation was based on a 5$\sigma$ level detection. Thus, although our reported diameter matches those from other works, it is likely a lower limit. Our $p_{V}$ of 25\% agrees with the 23\% reported by \citet{Geem2023} and the 22\% by \citet{Fornasier2024} and match typical albedos observed for S-complex asteroids \citep{DOMINGUE2002205, Mainzer2011_class, Thomas2011}. Our higher $p_{V}$ solution of 30\% results from the lower JPL-$H_{V}$ value. Since \citet{Fornasier2024} reported an $H_{V}$ of 18.94$\pm$0.05, similar to our MOC solution, we suggest that a $p_{V}$ of 25\% is favored.

\subsubsection{Mission Support Observations: Didymos-Dimorphos post-DART Impact Ejecta Characterization}

The Double Asteroid Redirection Test (DART) Planetary Defense spacecraft successfully impacted and modified the orbit of Dimorphos \citep{Thomas2023}, the satellite of the (65803) Didymos asteroid, on 2022 September 26 at 23:14 UTC \citep{Daly_2023}. Rapidly after the impact, the Didymos system was covered by an ejecta cloud excavated from the surface of Dimorphos \citep{Li_2023, Opitom2023, Murphy2023, Kareta_2023}. Understanding the properties and evolution of the ejecta is key to interpreting the consequences and efficiency of the DART impact. Thus, we observed the Didymos-Dimorphos system twice, at 11 hours and 9 days after the DART impact, in an effort to provide ejecta characterization and mission support studies. 

On 2023 September 27 (10 UTC) and October 6 (13 UTC), we acquired 8.7, 10.5, and 11.7~${\mu}$m observations with the MIRSI system. On September 27, we obtained Didymos fluxes of 699.1$\pm$85.7, 1791.7$\pm$36.6, and 2211.4$\pm$65.0~mJy at 8.7, 10.5, and 11.7~${\mu}$m, respectively. On October 6, we obtained Didymos fluxes of 639.1$\pm$64.4 989.0$\pm$35.6, and 880.0$\pm$61.0~mJy at 8.7, 10.5, and 11.7~${\mu}$m, respectively. 

\begin{figure}[h!]
    \centering
    \includegraphics[width=0.48\textwidth]{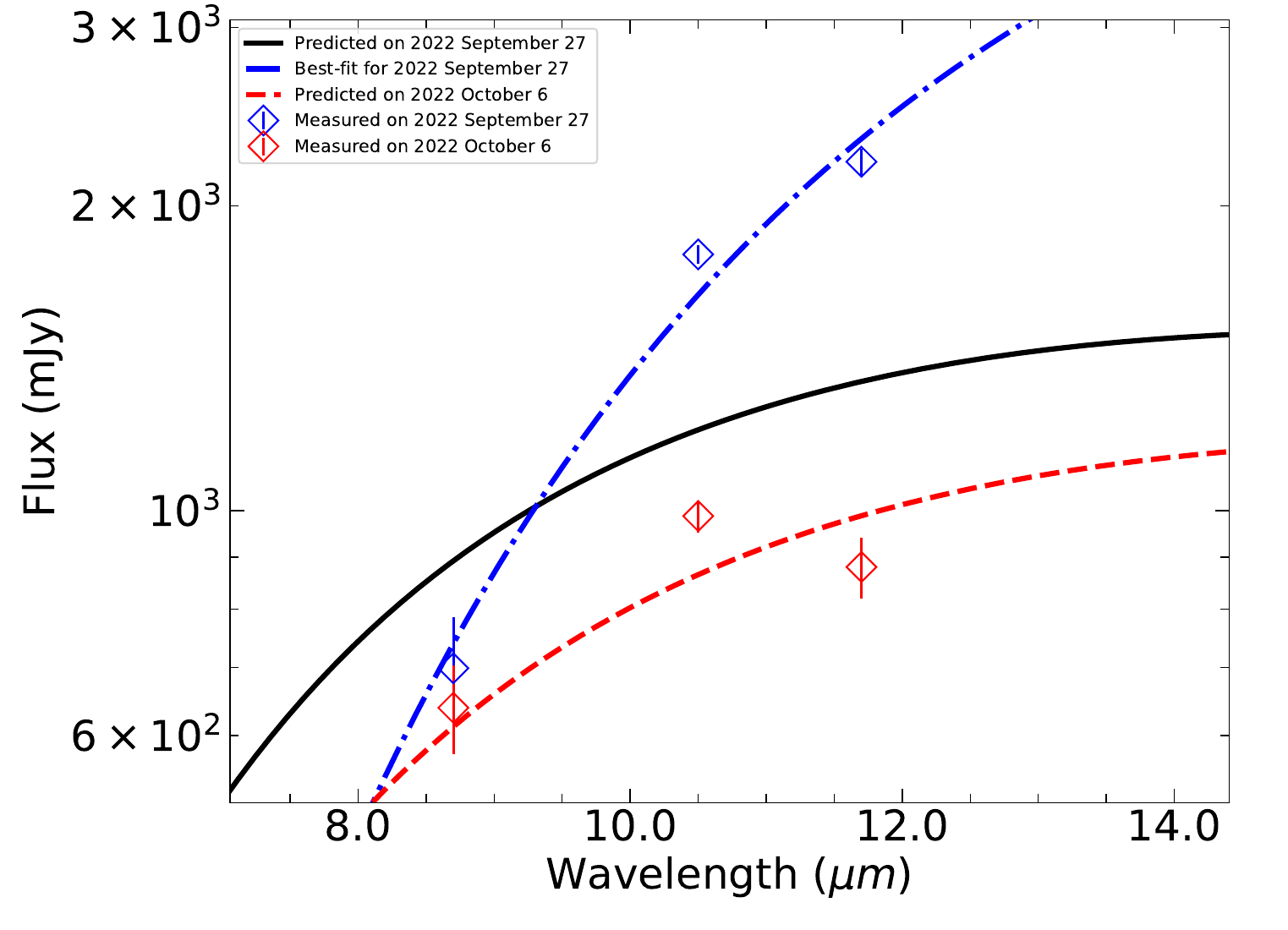}
    \caption{Predicted and measured spectral energy distribution of the Didymos-Dimorphos system. The black (solid) and red (dashed) lines illustrate the predicted spectral energy distribution (i.e., with no impact or dust) for 2022 September 27 and 2022 October 6, respectively. The blue and red diamonds represent the measured spectral energy distribution at 8.7, 10.5, and 11.7~$\mu$m for September 27 and October 6, respectively. The dash-dotted blue line shows the best-fit SED for the Didymos-Dimorphos system 11 hours after the DART impact.}
    \label{fig:didymos}
\end{figure}

In Figure \ref{fig:didymos}, we show the measured and the predicted SED for the Didymos system for the two observation epochs. The measured SED for the Didymos system 11~hours after the DART impact shows significant deviations from the predicted one. The predicted SEDs (black and red lines) in Figure~\ref{fig:didymos} assumed a combined cross-sectional diameter of 0.73~km for the Didymos system \citep{Daly_2024}, a geometric albedo of 17\% \citep{Rozitis2024}, and the respective geometries of Didymos at the time of the observations (see Table~\ref{tab:Table1}). 

The excess in measured flux (black diamonds in Figure \ref{fig:didymos}) at 10.5 and 11.7~${\mu}$m on September 27 directly reflects the properties of the ejecta being optically thick at these or even at shorter wavelengths \citep{Lazzarin2023}. In addition, the steeper slope of the measured SED 11~hours after the DART impact suggests that the ejecta dust has a much lower blackbody temperature than predicted (T$\sim$315~K) for the Didymos-Dimorphos system. We used a similar approach for debris disk characterization around stars \citep{Su2005, Trilling2008} to obtain mass estimations for the Dimorphos ejecta from the observed excess in the thermal infrared flux. 

For the September 27 observations, we obtained that a blackbody with T$=$195~K and sphere diameter of $D_{eff}=3.31^{+0.78}_{-1.04}$~km best fit the measured SED. We evaluated the equivalent mass of a sphere with such diameter for the number of 10~${\mu}$m particles. We obtained an ejecta mass of $2.1-9.1{\times}10^{6}$~kg assuming a density range associated with ordinary chondrites and the Didymos system \citep{Britt2003, DUNN2013273} of 3500~kg~m$^{-3}$.

The Hubble Space Telescope ejecta characterization determined a particle size distribution from 1~$\mu$m to 5~cm dominated the ejecta \citep{Moreno_2023} with a broken power law at 3~mm. Within this particle range, \citet{Moreno_2023} estimated the dust mass ejecta to be around 0.9-6${\times}10^{6}$~kg. Thus, our mass ejecta estimations agree with those from \citet{Moreno_2023}. As expected, this is a lower estimate as we were sensitive to particles between 8 to 11~$\mu$m in size. More realistic measurements should contemplate the presence of mm particles to meter-sized boulders in the ejecta \citep{Kareta_2023, Roth2023, Moreno_2023}, which is expected to dominate the mass of the ejecta. Moreover, millimeter-sized ejecta characterization from \citet{Roth2023} determined a mass of 1.3-6.4${\times}10^{7}$~kg. Merging estimates from this work, \citet{Moreno_2023}, and \citet{Roth2023}, covering from micrometer to millimeter-size particles, the minimum total ejecta mass is rounding the $\sim$7${\times}10^{7}$~kg or about 1.6\% of Dimorphos's mass.

Another interesting outcome of our ejecta characterization arises from the October 6 observation. Notably, the measured SED (see Figure \ref{fig:didymos}) 9 days after the impact lack of excess in measured thermal fluxes at 10 and 11~${\mu}$m suggesting that many of these particles left the system to the interplanetary space likely by solar radiation pressure \citep{Rossi_2022, Moreno_2023} or reaccumulated into the surfaces of Didymos and Dimorphos. Additionally, thermal modeling analysis yielded best-fit diameter of 0.749$_{-0.072}^{+0.247}$~km and $p_{V}$ of 0.17$_{-0.12}^{+0.02}$ which agrees with those measurements from previous works \citep{Rozitis2024, Daly_2024}.

\subsubsection{Rapid-response of Recently-discovered NEOs \label{sec:recent-discovered}}

The MIRSI system will play a crucial role in future planetary defense characterizations. One of the primary goals in the original proposal to NASA to update MIRSI was to change the cryogenic system so that the instrument would be constantly cold and ready to be used. The closed-cycle refrigerator update in MIRSI helps keep the camera between 5 to 10~K without the need for liquid helium or nitrogen. Thus, the camera is always in operation mode to perform observations. This provides a unique advantage for observing recently discovered (small) NEOs or performing last-minute high-priority observation of a potential impactor whose size and albedo must be determined.

The first participation of MIRSI into planetary defense, after science commissioning started, was on 2023 March 25, with the rapid response characterization of NEO 2023~DZ2. This effort was made in contribution to the fourth planetary defense exercise by the International Asteroid Warning Network (IAWN) \citep{Vishnu2024}. In this program, we performed two additional rapid response observations of two recently discovered NEOs as a planetary defense exercise, applying lessons learned from the 2023~DZ2 campaign. 

In planetary defense practices, determining the potential impactor taxonomy, size, and albedo is key for predicting the conditional damage risks \citep{Reddy2022, Vishnu2024}. On 2023 April 6, we observed NEO 2023~GM six days after discovery when the asteroid was at $V_{mag}=$16.9~mags and at 0.02~AU from Earth. We measured a diameter of 72$\pm$5~meters and a $p_{V}$ of 0.12$_{-0.05}^{+0.24}$ with our optically derived $H_{V}=$23.67$\pm$0.18~mags. While the taxonomy of this asteroid is unknown, we suggest either an S- or C-complex given that the albedo solution falls within the expected one for those objects \citep{DOMINGUE2002205, Mainzer2011_class}. Additionally, our simultaneous MOC photometry suggests a rotation period of 3$\pm$1~minutes. Such a rapid rotation period and taxonomy suggest that 2023 GM is likely a monolith or a non-zero cohesion strength asteroid \citep{SCHEERES2018183, BRISSET2022105533} of about 70 meters in diameter. 

On 2024 February 14, we observed NEO 2024~CG2 six days after discovery when the asteroid was at $V_{mag}=17.24$~mags and at 0.03~AU from Earth. We measured an average diameter of 53$\pm$10~meters and a $p_{V}$ of 0.43$_{-0.09}^{+0.10}$ with our optically derived $H_{V}=22.94$$\pm$0.15~mags. Based on this range of albedos, we suggest that 2024~CG2 is likely an X-complex as those objects usually show albedos higher than 30\% \citep{Thomas2011}.

\subsubsection{Non-detection with MIRSI \label{sec:non-detection}}

\begin{figure}[h!]
    \centering
    \includegraphics[width=0.47\textwidth]{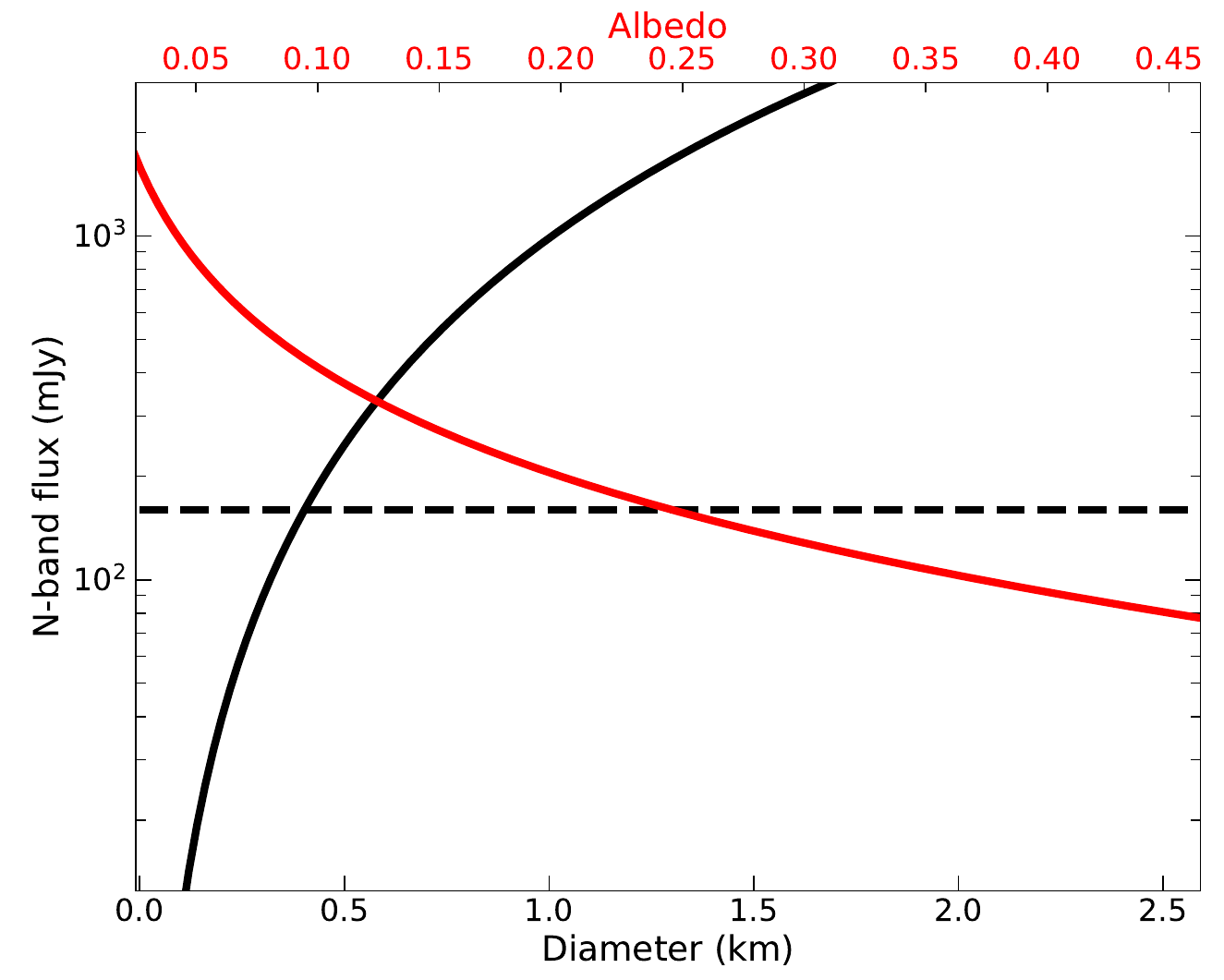}
    \caption{Predicted N-band fluxes for 1989 ML for a range of albedos (red line) and diameters (black line). The horizontal dashed line corresponds to the 5$\sigma$ flux we would have detected in a $\sim$47~minutes of integration time. The red line flux solutions were computed for a sphere with radius obtained from the \citet{Fowler1992} relationship and assuming the JPL-Horizons $H_{V}$ solution. The black line flux solutions were computed for various sphere sizes with a fixed $p_{V}$ of 50\%. We conclude that this object must have an albedo higher than 0.25 and a diameter smaller than 400~meters.}
    \label{fig:1989ML}
\end{figure}

In Table \ref{tab:Table2}, we present a few observed NEOs for which we could not obtain a strong detection (i.e., $>10~{\sigma}$). Although these NEOs were poorly detected, several conservative constraints in the target diameter and $p_{V}$ can be made. Below, we discuss the range of $p_{V}$ and the diameter solution that can be established for these NEOs based on the amount of telescope wall-clock integration time we spent to detect a given source and predicted synthetic N-band fluxes. We clarify that we treat a non-detection as those measurements at the $1~{\sigma}$ level, and which visual identification in the mosaics such as in Figure \ref{fig:7482_mosaic} was not possible. In Figure~\ref{fig:1989ML}, we show a visualization of the approach implemented in this section.

\textit{(3103) Eger 1982~BB}. We spent 64~minutes on Eger and should have detected a $\sim$280~mJy source. The predicted flux of Eger with a 15\% albedo for the given geometry we observed yielded an N-band flux of $\sim$360~mJy. We obtained this prediction by assuming a sphere size extrapolated from the JPL-Horizons $H_{V}$ solution and the \citep{Harris2002} relationship. We infer that Eger's $p_{V}$ should be higher than 20\% and its diameter lower than 2~km. Otherwise, a lower albedo and/or larger diameter would have yielded a stronger detection in the observed time. Furthermore, Eger is known as an Xe-type asteroid \citep{Binzel2019}, which usually presents albedos higher than $\sim$30\% \citep{Thomas2011}. In addition, a higher albedo solution and $D<2.1$~km for Eger is confirmed by previous measurements from Spitzer ExploreNEOs, which reported a $p_{V}$ of 42\% and $D$ of 1.5~km \citep{Trilling_2010}. \citet{Usui2011} measured an even larger $p_{V}$ of 0.81 and $D$ of 1.2~km. Additionally, \citet{Alilagoa2018} reported a $D$ of 1.65~km and a $p_{V}$ of 0.68. Consolidating the literature measurements, it seems likely that Eger is a high albedo object with a diameter lower than 2~km, as suggested by our estimates.

\textit{(138127) 2000~EE14}. We spent 55~minutes on NEO 2000~EE14, and we should have detected a $\sim$300~mJy source at 10$\sigma$ level or a $\sim$150~mJy at 5~$\sigma$ level. The predicted thermal flux with H$_{JPL}$ and $p_{V}=15\%$ was around 140~mJy. Thus, our lack of detection presents the possibility that the object has a smaller diameter and/or higher albedo surface. Since we did not detect any flux above the 1$\sigma$ level, we predict that NEO 2000~EE14 is a high albedo object with $p_{V}>30\%$, and its diameter should be lower than 1.5~km. Indeed, its measured diameter of 0.72~km and albedo between 44 to 52\% \citep{Usui2011, Nugent_2015} confirm our established limits. We also point out that the variable weather conditions during the 2000~EE14 observation could slightly affect our null measurement. Moreover, our albedo and diameter range are conservative enough to capture the lower and upper limits due to the possibility of the weather affecting them.

\textit{(388188) 2006~DP14}. We observed 2006~DP14 with MIRSI for about 31~minutes of wall clock time and measured a 1$\sigma$ flux of 30~mJy. We argue that $p_{V}$ for 2006~DP14 should be higher than 25\% and $D<0.65~km$ since a lower albedo or larger diameter would have yielded a stronger detection. The optical albedo of 2006~DP14 is unknown, but our constraint of $p_{V}>25\%$ is reasonable \citep{DOMINGUE2002205, Mainzer2011_class, Thomas2011} for its known S-type taxonomy \citep{Binzel2019}.

\textit{(10302) 1989~ML}. We integrated for 47~minutes on 1989~ML and obtained a 1$\sigma$ detection of 22~$mJy$. We predicted the object to be around 270~$mJy$, which would have produced an 8$\sigma$ detection in $\sim$47~minutes. (See Figure~\ref{fig:1989ML} for a visual representation of the predicted fluxes of 1989 ML for a range of albedo and diameters). Thus, we place a conservative lower limit on $p_{V}$ of 25\% and $D<0.4~km$. Our established limits fall within \citet{Mueller_2011} estimates, which derived a $p_{V}$ of 47\% and a diameter of 0.24~km.

\subsubsection{Main-Belt Observations: 23 Thalia}

The SED of Main-Belt asteroids reaches maxima around 20~${\mu}$m. Although MIRSI currently lacks a 20~${\mu}$m entrance window for observations at these wavelengths or longer, many large Main-Belt asteroids become bright enough ($>$1~Jy) to be detected at $<$16~$\mu$m. As an example, on 2024 January 29, we observed 23 Thalia while the asteroid was 1.3~AU from Earth and with $V_{mag}=$10.36~mags. We measured fluxes of 24.658$\pm$1.233, 34.637$\pm$1.731, and 52.070$\pm$2.604~Jy at 8.7, 10.5, and 11.7~${\mu}$m. The thermal modeling analysis yielded a best-fit $D=124.2^{+43.6}_{-23.5}$~km, $p_V=0.15^{+0.05}_{-0.04}$, and $\eta=$1.06$^{+0.44}_{-0.30}$. Our estimated diameter is consistent within uncertainties with the range of 95 to 120~km suggested by previous studies \citep{Usui2011, Hanus2013, Masiero2011}. Our $p_{V}$ solution is also consistent with the 15\% to 31\% range found by previous works \citet{Usui2011, Masiero2011, Nugent2016} and well match the S-type taxonomy of 23 Thalia \citep{Lazzaro2004}.


\section{Conclusion \& Future Work}

In this work, we demonstrated and highlighted the capability of MIRSI, the newly refurbished mid-infrared camera in the NASA IRTF, by providing diameter and albedo measurements for 31 NEOs. Our survey presented the first albedo and diameter solution for eleven NEOs whose properties were previously unknown. For most of our observations, we were able to simultaneously constrain their absolute magnitude at the time of the thermal observation. We successfully provided mission support observations for the JAXA-Haybusa2$\sharp$ project by measuring the albedo and diameter of the NEO (98943) 2001 CC21, which the spacecraft will fly-by in 2026. We also provided ejecta characterization for the NASA/DART Dimorphos impact by providing an ejected mass estimate and implications for the time evolution of $\sim$10~$\mu$m particles in the ejecta. Our observations of three recently discovered NEOs highlight the effectiveness of an easily accessible and operable instrument to enable rapid-response characterization at the NASA IRTF in support of planetary defense efforts. 

This NEO observational program with MIRSI will continue providing albedo and diameter measurements and mission support characterization for targets of interest. As additional future plans, we will contemplate several ideas listed below:

\begin{enumerate}
    \item There are ongoing strategies led by the NASA-IRTF staff to improve the camera performance after the science array detector upgrade in January 2024. Thus, we expect the camera performance described in \citet{Hora2024} to improve by a factor of 4 to 10.
    \item We will improve the simultaneous optical photometry characterization by combining the Opihi\footnote{\url{https://irtfweb.ifa.hawaii.edu/~opihi/}} camera to cross-correlate it with MOC. This will provide better optical photometry calibration and, by extension, obtain more robust $H_{V}$ measurements. 
    \item We will extend our survey with a focus on very small ($D<150$~meters) recently discovered NEOs in order to better estimate the size-frequency and albedo distribution of the small NEO population. In addition, MIRSI will play an important role in future Target of Opportunity Observations (ToOs) and planetary defense efforts such as in \citet{Reddy_2024}. 
    \item MIRSI will play an important role in the thermal characterization of the potentially hazardous asteroid 99942 Apophis before, during, and after the encounter in 2029. In addition, MIRSI will continue to serve as an instrument to provide further mission support observations, such as those done in this program for JAXA-Haybusa2$\sharp$ and NASA-DART.
    \item We will test the MIRSI grism mode to perform spectral characterization of NEOs and main-belt asteroids once the performance of the camera improves.
\end{enumerate}

\software{Astropy \citep{AstroPy2013},
          Scipy \citep{scipy2020},
          Numpy \citep{numpy2020},
          Matplotlib \citep{matplotlib2007}
    }

\acknowledgements 
\ This material is in part supported by the National
Science Foundation Graduate Research Fellowship Program under grant No.\ 2021318193 to ALO. Any opinions, findings, conclusions, or recommendations expressed in this material are those of the author(s) and do not necessarily reflect the views of the National Science Foundation.
This work was partially funded by a grant from the NASA Solar System Observations/NEOO program (NNX15AF81G). The original MIRSI instrument was funded by NSF grant 9876656 and support from Boston University.
Part of the computational analyses were carried out on Northern Arizona University's Monsoon computing cluster, funded by Arizona's Technology and Research Initiative Fund.
The Infrared Telescope Facility is operated by the University of Hawaii under contract 80HQTR19D0030 with the National Aeronautics and Space Administration.
We thank all the NASA-IRTF staff for their support and incredible efforts on MIRSI, especially Charles Lockhart, Mike Connelley, Mike Kelii, and John Rayner.

\bibliographystyle{aasjournal}
\bibliography{references}{}

\end{document}